\documentclass[acmsmall, screen, nonacm]{acmart}

\usepackage{graphicx}
\usepackage{xspace}
\usepackage{tikz}
\usepackage{xcolor}
\usepackage{subcaption}
\newcommand*\circled[1]{\tikz[baseline=(char.base)]{
            \node[shape=circle,fill,inner sep=1pt] (char) {\textcolor{white}{#1}};}}
\usepackage{multirow}
\usepackage{array}
\usepackage{tabularx}
\usepackage{seqsplit}
\usepackage{booktabs}  
\usepackage{listings}
\usepackage{caption}
\usepackage{xspace}
\usepackage{tcolorbox}
\usepackage{amsmath}
\usepackage{array}
\usepackage{fix-cm}
\usepackage{tikz}
\usepackage{amsmath}
\usetikzlibrary{shapes.callouts, shadows}
\usepackage{xcolor}
\usepackage{color, colortbl,booktabs}
\definecolor{Gray}{gray}{0.9}

\usepackage{listings}
\usepackage{tikz}

\newcommand{\peter}[1]{\textcolor{red}{{\it [Peter says: #1]}}}

\newcommand{\toolwocc}{\textit{DepGraph\textsubscript{w/o Code Change}}\xspace}
\newcommand{\tool}{\textit{DepGraph}\xspace}
\newcommand{\tooltable}{DepGraph\xspace}
\newcommand{\toolwocctable}{DepGraph\textsubscript{w/o Code Change}\xspace}

\newcommand{\graphname}{\textit{Dependency-Enhanced Coverage Graph}\xspace}
\newcommand{\graphnameLower}{\textit{dependency-enhanced coverage graph}\xspace}

\newcommand{\phead}[1]{\vspace{1mm} \noindent {\bf #1}}
\newcommand{\uhead}[1]{\vspace{1mm} \noindent {\underline {#1}}}

\newcommand{\rqboxc}[1]{\begin{tcolorbox}[left=4pt,right=4pt,top=4pt,bottom=4pt,colback=gray!35,colframe=gray!35,before skip=3pt,after skip=3pt]#1\end{tcolorbox}}

\definecolor{darkgreen}{rgb}{0.01, 0.75, 0.24}
\newcommand{\tinier}{\fontsize{5pt}{1pt}\selectfont}
\AtBeginDocument{%
  }

\setcopyright{acmcopyright}
\copyrightyear{2018}
\acmYear{2018}
\acmDOI{XXXXXXX.XXXXXXX}

\acmConference[International Conference on the Foundations of Software Engineering (FSE) 2024]{Make sure to enter the correct
  conference title from your rights confirmation email}{June 03--05,
  2018}{Woodstock, NY}
\acmPrice{15.00}
\acmISBN{978-1-4503-XXXX-X/18/06}

\begin{document}

\title[]{Towards Better Graph Neural Network-based Fault Localization Through Enhanced Code Representation}


\author{Md Nakhla Rafi}
\email{r_mdnakh@encs.concordia.ca}
\affiliation{%
  \institution{Concordia University}
  \city{Montréal}
  \state{Québec}
  \country{Canada}
}

\author{Dong Jae Kim}
\email{djaekim086@gmail.com}
\affiliation{%
  \institution{DePaul University}
  \city{Illinois}
  \state{Chicago}
  \country{United States}
}

\author{An Ran Chen}
\email{anran6@ualberta.ca}
\affiliation{%
  \institution{University of Alberta}
  \city{Edmonton}
  \state{Alberta}
  \country{Canada}
}

\author{Tse-Hsun (Peter) Chen}
\email{peterc@encs.concordia.ca}
\affiliation{%
  \institution{Concordia University}
  \city{Montréal}
  \state{Québec}
  \country{Canada}
}

\author{Shaowei Wang}
\email{Shaowei.Wang@umanitoba.ca}
\affiliation{%
  \institution{University of Manitoba}
  \city{Winnipeg}
  \state{Manitoba}
  \country{Canada}
}

\renewcommand{\shortauthors}{Rafi et al.}

\begin{abstract}
      Automatic software fault localization plays an important role in software quality assurance by pinpointing faulty locations for easier debugging. Coverage-based fault localization is a commonly used technique, which applies statistics on coverage spectra to rank faulty code based on suspiciousness scores. However, statistics-based approaches based on formulae are often rigid, which calls for learning-based techniques. Amongst all, {\it Grace}, a graph-neural network (GNN) based technique has achieved state-of-the-art due to its capacity to preserve coverage spectra, i.e., test-to-source coverage relationships, as precise abstract syntax-enhanced graph representation, mitigating the limitation of other learning-based technique which compresses the feature representation. However, such representation is not scalable due to the increasing complexity of software, correlating with increasing coverage spectra and AST graph, making it challenging to extend, let alone train the graph neural network in practice. In this work, we proposed a new graph representation, {\it DepGraph}, that reduces the complexity of the graph representation by 70\% in node and edges by integrating interprocedural call graph in the graph representation of the code. Moreover, we integrate additional features -- code change information -- in the graph as attributes so the model can leverage rich historical project data. We evaluate \tool using Defects4j 2.0.0, and it outperforms {\it Grace} by locating 20\% more faults in Top-1 and improving the Mean First Rank (MFR) and the Mean Average Rank (MAR) by over 50\% while decreasing GPU memory usage by 44\% and training/inference time by 85\%. Additionally, in cross-project settings, {\it DepGraph} surpasses the state-of-the-art baseline with a 42\% higher Top-1 accuracy, and 68\% and 65\% improvement in MFR and MAR, respectively. Our study demonstrates \tool's robustness, achieving state-of-the-art accuracy and scalability for future extension and adoption.
\end{abstract}


\received{20 February 2007}
\received[revised]{12 March 2009}
\received[accepted]{5 June 2009}


\maketitle

\section{Introduction}

Locating and fixing software faults is a time-consuming and manual-intensive process. A prior study~\cite{hait2002economic} found that more than 70\% of the software development budgets are for software testing and debugging. 
As software projects continue to grow in complexity and scale, the need for efficient and accurate fault localization techniques further increases. 
Hence, to assist developers and reduce debugging costs, researchers have proposed various fault localization techniques~\cite{li2019deepfl, lou2021boosting, qian2023gnet4fl, li2021fault, abreu2009spectrum, sohn2017fluccs} to expedite the debugging process. 

Traditional fault localization techniques, such as spectrum-based fault localization (SBFL), analyze the coverage of the passing and failing test cases to identify the potential locations of the fault that triggers the test failure. SBFL techniques are based on the intuition that a code element that is covered by more failing and fewer passing test cases is more likely to be faulty. In past decades, researchers have proposed different SBFL techniques, such as \textit{Ochiai}~\cite{abreu2006evaluation}, by crafting various formulas to rank the code elements based on the test coverage results. However, one major limitation is that such formulas may not generalize well to various faults and projects \cite{wong2016survey,xie2013theoretical,le2013theory,zou2019empirical}.

Due to recent advances in machine learning and deep learning, there is a surge in learning-based fault localization techniques ~\cite{sohn2017fluccs, zhang2019empirical, li2021fault, li2017transforming, li2019deepfl, zhang2019cnn}. These techniques enhance SBFL's potential by training models to rank the likelihood of faulty code elements, which often provide better accuracy and generalizability compared to traditional techniques. Learning-based techniques can combine various metrics in addition to test coverage to improve fault localization accuracy. For instance, 
\textit{FLUCCS}~\cite{sohn2017fluccs} and \textit{DeepFL}~\cite{li2019deepfl} incorporate both test coverage and code structure metrics into training deep learning models for fault localization, and showed significant improvement over traditional SBFL techniques. 

In recent years, Graph Neural Networks (GNN)-based fault localization techniques have shown promising results. 
GNN-based techniques~\cite{lou2021boosting, qian2023gnet4fl} represent the code structure as a graph, in which the coverage information is represented as edges between different types of nodes (e.g., method nodes or line nodes). For example, as one of the earliest works, \textit{Grace}~\cite{lou2021boosting} uses test cases and nodes in the abstract syntax tree (AST) as nodes in the graph representation. \textit{Grace} constructs edges between test nodes and statement nodes to represent the dynamic coverage information in the graph. However, there are limitations in this graph representation. By only examining the test coverage and AST nodes, the graph representation is missing the caller-callee information. Moreover, the graph only contains structural code information, but 
as shown in previous studies~\cite{sohn2017fluccs, chen2022useful}, historical code evolution information can also be valuable for fault localization. 

In this paper, we proposed a novel GNN-based fault localization technique, \tool, which integrates interprocedural method calls and historical code evolution in the graph representation. 
In particular, \tool leverages a \graphname to enhance the code representation. We first constructed a unified graph representation based on the code structure, interprocedural method calls, and test coverage. Different from prior graph representations, \graphnameLower is able to eliminate edges among methods that do not have call dependencies; thus, reducing noise in the graph. We then add code churn as attributes to the nodes in the graph, providing historical code evolution information to the GNN. 

We evaluated \tool on the widely-used Defects4j (V2.0.0) benchmark \cite{just2014defects4j}, which contains 675 real-world faults from 14 open-source Java projects. Our results show that \tool outperforms both \textit{Grace} and \textit{DeepFL}, the state-of-the-art learning-based fault localization technique~\cite{lou2021boosting, li2019deepfl}. 
Compared to {\it Grace}, \toolwocc (equivalent to {\it Grace} + \graphnameLower) locates 13\% more faults at Top-1 and achieves over 40\% improvements in both Mean First Rank (MFR) and Mean Average Rank (MAR). 
By further integrating code change data (i.e., code churn and method modification count) as additional features for \tool, \tool can locate 23 additional faults within Top-1 compared to \toolwocc.
This underscores the significance of refining graph representation and combining it with information from software development history. Furthermore, we evaluated the computing resources that can be reduced by adopting the \graphnameLower. Our results show that \graphname dramatically reduces the graph's size by 70\% and minimizes GPU memory consumption by 44\%, showing potential directions on adopting GNN-based techniques on larger projects. Finally, in cross-project settings, \tool significantly surpasses {\it Grace}, demonstrating a 42\% improvement in Top-1 accuracy. 

The paper makes the following contributions: 
\begin{itemize}
    \item We proposed a novel GNN-based technique, \tool, which incorporates call dependency and code evolution information in the graph representation of a project. Our findings show that \tool improves Top-1, MFR, and MAR by 20\%, 55\%, and 52\%, respectively, compared to {\it Grace}~\cite{lou2021boosting}.
    
    \item Compared to \toolwocc, adding code change information further improve Top-1 by 7\%. Future studies should consider integrating additional information into the graph to improve fault localization results. 

    \item Adopting the \graphname helps reduce both GPU memory usage (from 143GB to 80GB) and training/inference (from 9 days to 1.5 days). Future studies should consider compacting the graph representation to reduce the needed resources to train and run GNN-based FL techniques. 

    \item Adopting \graphnameLower and code change information can help locate 10\% to 26\% additional faults compared to \textit{Grace}, without missing the faults that \textit{Grace} could locate. We also find that the additional faults that \tool can locate are related to the method interactions, loop structures, and call relationships, which further shows the importance of our \graphnameLower.  

    \item In cross-project settings, \tool and \toolwocc, significantly outperform {\it Grace}, with \toolwocc showing a 23\% increase in Top-1 accuracy and \tool achieving a even higher improvement at 42\%. \tool trained in cross-project settings has similar or even better results compared to the baseline techniques that are trained using data from the same project. 
\end{itemize}

\noindent {\bf Paper Organization.} Section~\ref{sec:related} discusses related work. Section~\ref{motivational_example} provides a motivating example. Section~\ref{section:approach} describes our technique, \tool. Section~\ref{section:result} presents the experiment results. Section~\ref{sec:threat} discusses the threats to validity. Section~\ref{sec:conclusion} concludes the paper.

\section{Related Work}
\label{sec:related}




\phead{Spectrum-based fault localization.} Spectrum-based fault localization (SBFL)~\cite{abreu2006evaluation,jones2002visualization,wong2013dstar,abreu2009spectrum} leverages statistical formulas to compute the suspiciousness of each code element (e.g., method) based on the test results and program execution. The intuition behind SBFL is that the code elements covered by more failing tests and fewer passing tests are more likely to be faulty. 
While SBFL has been widely studied, their effectiveness is still limited in practice~\cite{kochhar2016practitioners,xie2016revisit}. 
Several prior studies~\cite{cui2020improving, wen2019historical, chen2022useful, xu2020every} have proposed leveraging new information, such as code changes~\cite{wen2019historical,chen2022useful} or mutation information~\cite{cui2020improving,xu2020every}, to improve the accuracy of SBFL. However, the computed suspiciousness scores still heavily rely on code coverage and may be generalized well to other faults or systems. 

\phead{Learning-based fault localization.}
Recently, there has been extensive research effort on leveraging learning-based methods to enhance the capabilities of SBFL~\cite{sohn2017fluccs, zhang2019empirical, li2021fault, li2017transforming, li2019deepfl, zhang2019cnn}. These techniques learn and derive the suspiciousness scores by learning from historical faults. Researchers have proposed using various machine learning techniques for fault localization, such as using radial basis function networks \cite{wong2011effective}, back-propagation neural networks \cite{wong2009bp}, multi-layer perceptron neural networks \cite{zhang2019cnn}, and convolutional neural networks \cite{zhang2019cnn, li2021fault, albawi2017understanding}. Some learning-based FL techniques integrate the suspiciousness scores computed by existing SBFL approaches with other relevant metrics. For instance, \textit{FLUCCS} \cite{sohn2017fluccs} combines SBFL-derived scores with other metrics, like code complexity and code history data. \textit{CombineFL} \cite{zou2019empirical} and \textit{DeepFL} \cite{li2019deepfl} combine features from diverse dimensions, including spectrum-based, mutation-based \cite{moon2014ask, papadakis2015metallaxis, dutta2021msfl}, and information retrieval \cite{zheng2016fault} scores to enhance fault localization accuracy. Le et al. \cite{le2015information} implemented a multi-modal technique that integrates information retrieval with program spectra for fault localization.

Due to recent advances in graph neural networks (GNNs), researchers have proposed using graphs to represent source code for learning-based fault localization~\cite{qian2023gnet4fl, lou2021boosting, qian2021agfl, xu2020defect}.  
Xu et al. \cite{xu2020defect} apply GNNs to capture the source code context, representing fault subtrees as directed acyclic graphs for detect prediction. 
\textit{AGFL}~\cite{qian2021agfl} employs vector transformations of the abstract syntax tree (AST) nodes to represent structural code information as graphs. 
\textit{Grace}~\cite{lou2021boosting} constructs the nodes and edges in the graph using test cases and code nodes from the program. Then, \textit{Grace} combines the constructed graph with test coverage, adding dynamic test execution information to the graph.  
\textit{GNET4FL}~\cite{qian2023gnet4fl} follows a similar code representation but uses the test results directly as node attributes in the AST nodes and uses GraphSAGE, a more advanced GNN architecture. 
\textit{GMBFL}~\cite{wu2023gmbfl} combines graph representation learning with Mutation-Based Fault Localization methods, utilizing a Gated Graph Attention Neural Network (GGANN) to extract features from the graph for identifying faulty program entities. 

Although GNN-based FL techniques have shown promising results compared to other FL techniques, there are some limitations in prior studies. The graph representation in prior studies may not accurately represent the code structure, and the information in the graph is only constrained to using code structure or test execution. 
Hence, in this paper, we propose \tool, which integrates interprocedural call dependency graph to enhance the code representation and adds software evolution information (i.e., code changes) to the nodes in the graph. 
Our findings show that \tool can improve the state-of-the-art GNN-based techniques (i.e., \textit{Grace}) by over 50\% in mean average rank and mean first rank, 
as well as improving the accuracy of identifying the most relevant faults at the top.
In addition, our graph representation can reduce the computing resources by over 44\% (total GPU memory reduced from 143GB to 80GB) and the training time by over 80\% (from 9 days to 1.5 days). Our findings open up potential future directions and highlight the importance of code representation and adding additional information to the graph.

\section{Motivation}\label{motivational_example}
In this section, we highlight the limitations of an existing graph representation of the code employed by existing GNN-based FL techniques~\cite{lou2021boosting, qian2023gnet4fl}. 
In \textit{Grace}~\cite{lou2021boosting}, the code is first represented as many small graphs, where the root nodes are the {\sf MethodDeclaration} nodes in the abstract syntax tree (AST) for both source code methods and test cases. Other nodes in a graph are the remaining AST nodes in the method/test case, such as {\sf IfStatement} or {\sf ReturnStatement}, connected by edges that represent the dependencies of the AST nodes.
The graph is then integrated with the test coverage result, where the root nodes (i.e., methods or test cases) are linked to form a larger graph if they have a coverage relationship in test execution. 
Using such a code representation for training GNN models overcomes the limitations associated with existing fault localization approach in several ways: (1) It overcomes rigid formula-based (e.g., \textit{Ochiai}) SBFL technique by automatically learning to localize faults~\cite{zou2019empirical}, (2) it overcomes some learning-based techniques like \textit{DeepFL}~\cite{li2019deepfl} that pre-processes rich coverage information into single vector, which loses topological coverage relationships~\cite{lou2021boosting}, and finally, (3) it also preserves rich AST information of the code, thereby preserving the structural integrity of the code.


\begin{figure*}
    \centering
    \begin{subfigure}{0.25\columnwidth}
        \includegraphics[width=0.8\linewidth]{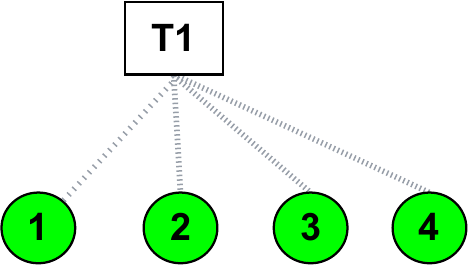}
         \caption{Coverage Representation}
    \end{subfigure}
    \hspace{1cm}
    \begin{subfigure}{0.25\columnwidth}
        \includegraphics[width=0.8\linewidth]{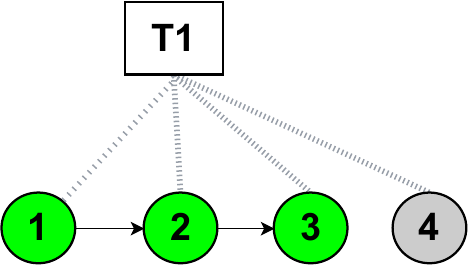}
         \caption{Call Graph Representation}
    \end{subfigure}
    \begin{tabular}{cc}

      \begin{lstlisting}
(*@\fbox{\textbf{\textit{TEST CODE}}}@*)
@Test
public void testNumberOverFlow() throws Exception(){
    doTestUnescapeEntity("&#12345678;", "&#12345678;") 
}

private void doTestUnescapeEntity(String, String)
throws Exception{
    assertEquals(expected, (*@entities.unescape(entity)); \circled{1}@*)
}
    
\end{lstlisting}   & 

  \begin{lstlisting}
(*@\fbox{\textbf{\textit{SOURCE CODE}}}@*)
class Entities {
    EntityMap map = (*@new Entities.LookupEntityMap(); \circled{2}@*)
    
    static class LookupEntityMap extends PrimitiveEntityMap (*@\circled{3}@*)

    static class PrimitiveEntityMap extends EntityMap{
        (*@{IntHashMap mapValueToName = new IntHashMap();} \circled{4}@*)
    }
}
    
\end{lstlisting} 
\\
    \end{tabular}
    \caption{Comparing graph representation for coverage and call graph for Lang-62. The green nodes correspond to the code statements (1--4). \emph{\textbf{(a)}} shows a coverage graph linking tests to source code statements without considering the method-level call graph. \emph{\textbf{(b)}} considers the method-level call graph, eliminating node 4 since it was not reachable in the call graph.}
    \vspace{-0.3cm}
    \label{fig:motive}
\end{figure*}



While the graph representation in \textit{Grace} has achieved state-of-the-art results, there are still several challenges with the current approach. We use a real bug, Lang-62, from the widely-used benchmark Defects4J (V2.0.0)~\cite{just2014defects4j} for illustration. Figure~\ref{fig:motive} shows Lang-62's test code and source code, and their corresponding coverage and call graph representations. Firstly, as shown in Figure~\ref{fig:motive}, the graph representation of code coverage only depicts the test-to-source method edge relationship, lacking the source-to-source edge relationship (i.e., the call graph relationship among source code methods). This omission may result in the loss of critical information about fault propagation in the source code, which could be helpful for enhancing fault localization results. Secondly, representing the entire code coverage as a graph may introduce inaccurate information into the candidate list for fault localization. 
For example, as shown in Figure~\ref{fig:motive}, while the test case \textit{``testNumberOverflow''} covers all source code entities (1-4), if we look at the source code, there is an absence of call relationship from (3) to (4). 
Although (4) is covered and an {\it IntHashMap} object is initiated, there is no evidence that the object is invoked during execution.  
Consequently, code elements like \textit{``IntHashMap''} might be incorrectly included in the graph, even if they are not pivotal to the fault\footnote{We provide the details of such statistics in our online repository~\cite{AnonymousSubmission9}}.

The motivating example shows that test coverage alone does not accurately portray the actual method calls within a program. Including such nodes and edges in the graph representation will increase complexity and noise in the graph representation learning process, making the GNN model less accurate and longer to train. Moreover, in addition to the structural information in the code, prior studies have found that historical information (e.g., code churn)~\cite{chen2022useful,sohn2017fluccs,wen2019historical} has a statistically significant relationship with faults. Hence, incorporating code change metrics may further enrich the graph representation by providing the evolution aspect of the code. 
We hypothesized that constructing the coverage graph based on test-to-source method relationships derived from the static call graph and incorporating code evolution information in the graph can further enhance GNN-based fault localization results. Below, we discuss our approach, \tool, in detail.

\begin{figure}
    \centering 
    \includegraphics[width=\textwidth]{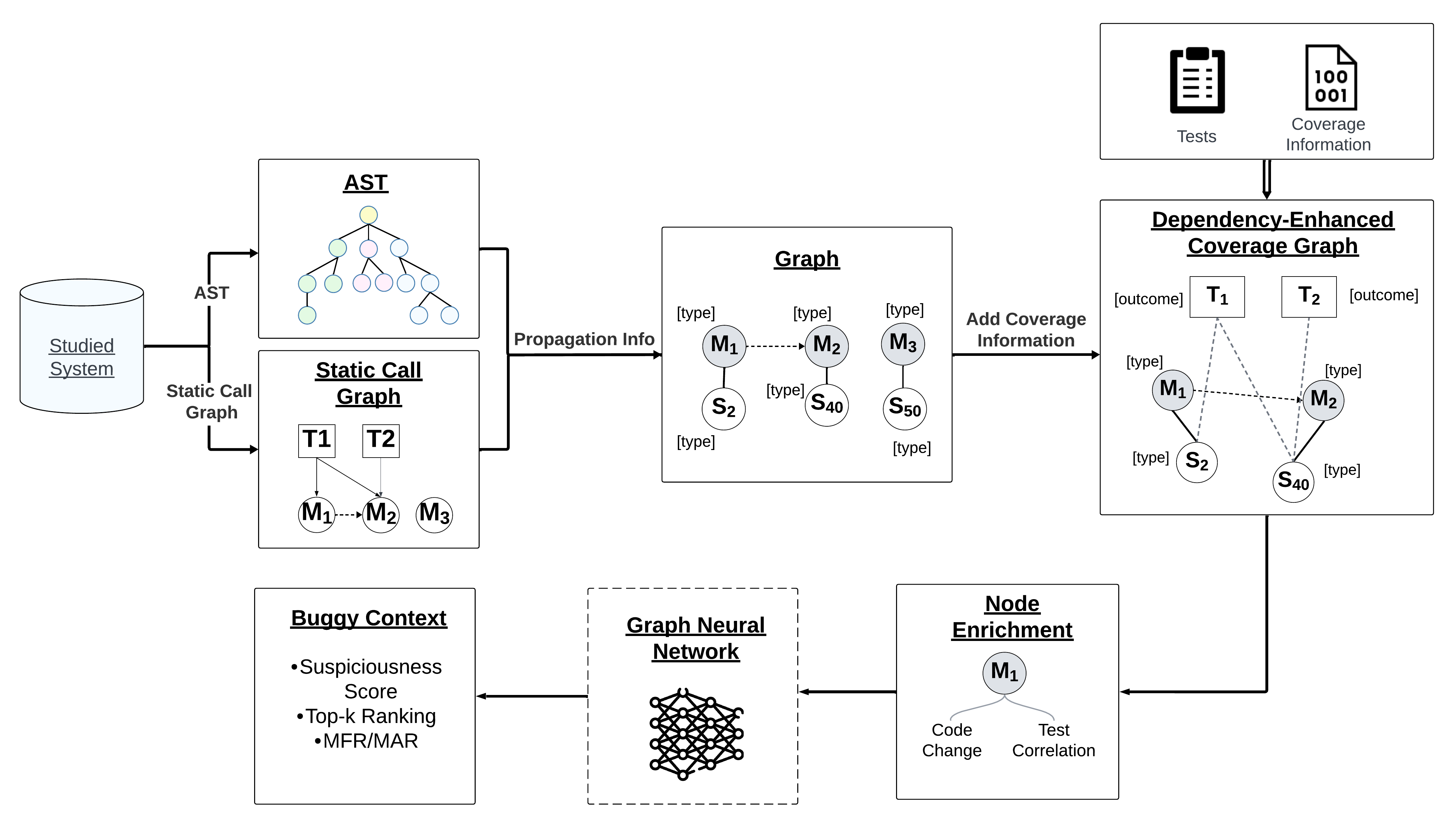}
    \caption{Overview of \tool. The term \( \text{type} \) denotes the AST node's type (e.g., M1 is a MethodDeclaration, and S2 is an IfStatement). Tests, such as T1, include test outcomes (e.g., pass or fail).}
    \label{fig:overall}
     \vspace{-0.3cm}
\end{figure}

\section{Approach}\label{section:approach}



We present \tool, a novel fault localization technique based on graph neural networks. Figure~\ref{fig:overall} presents an overview of \tool.
We implement \tool and conduct the experiment at the method-level by following prior studies~\cite{lou2021boosting, li2019deepfl, qian2023gnet4fl}, since debugging faults at the class level lack precision for effective localization~\cite{kochhar2016practitioners} and pinpointing at the statement level can be overly detailed and miss important context~\cite{parnin2011automated}.
Given the source and test code of a project, \tool constructs the abstract syntax tree (AST) for all the source methods. Then, \tool constructs an inter-procedural call graph to capture the relationships between different method calls.
The AST and the call graph are then merged to form a graph representation of the code. 
After getting the code coverage, we integrate the information on the covered code statement into the graph, pruning the AST vertices and the related methods that are not covered in the tests. 
To enhance the information in the graph and better identify faulty methods, we further integrate additional information (e.g., code churn) into the graph. Finally, the graph is fed to a Gated Graph Neural Network (GNN) model for training to locate potential faulty methods. Below, we discuss \tool in detail. 

\subsection{Dependency-Enhanced Coverage Graph}
\label{sec:graphRepresentation}
\begin{figure}
    \centering
    \includegraphics[width=.8\textwidth]{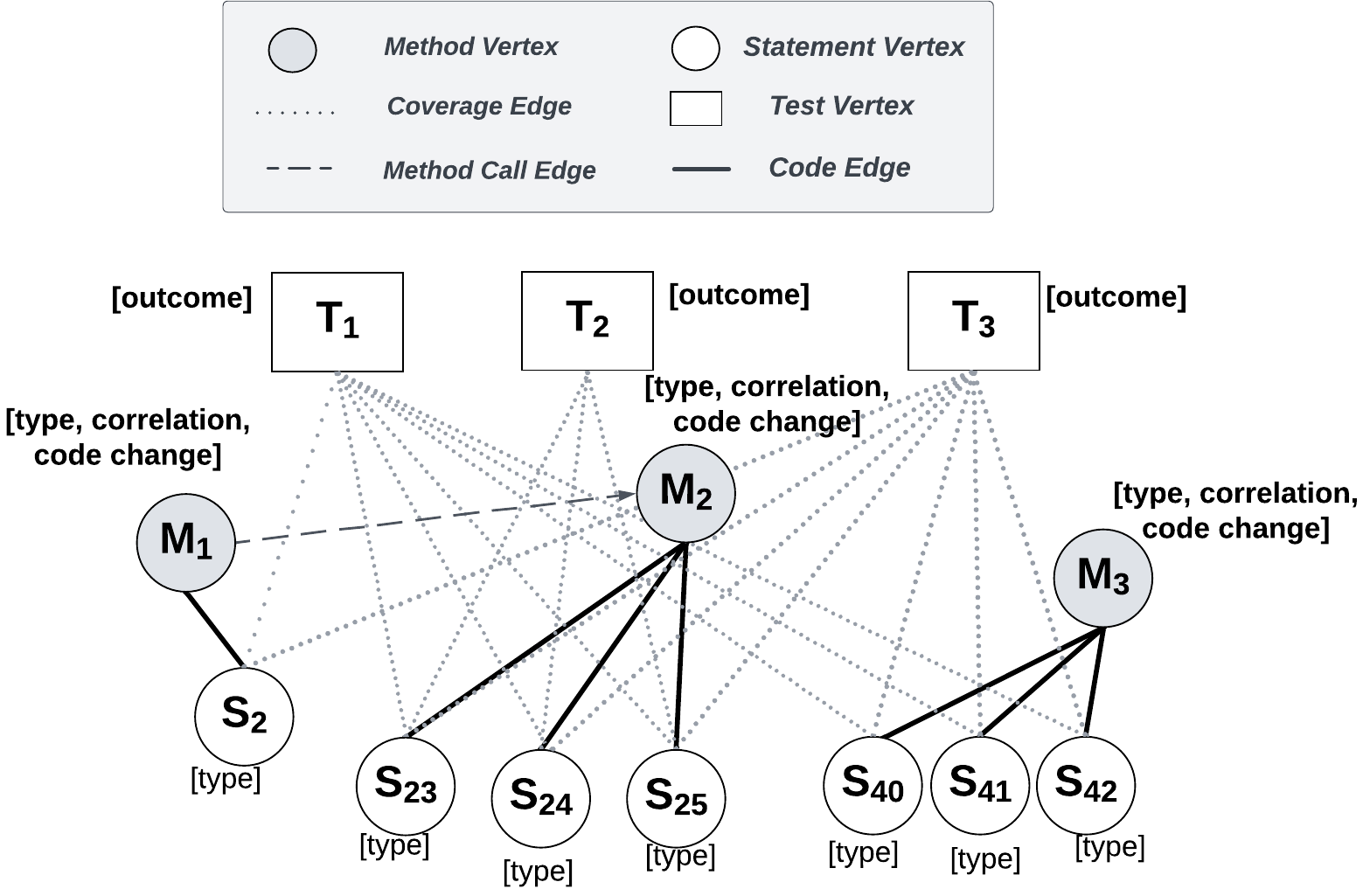}
    \vspace{-0.3cm}
    \caption{An example of the \graphname representation in \tool.} 
    \label{fig:full_graph}
    \vspace{-0.3cm}
\end{figure}

We represent the code, its call graph, and code coverage as a \graphname. 
As illustrated in Figure~\ref{fig:full_graph}, we represent the source methods, statements, and individual test methods as vertices (also referred to as nodes). The nodes are then connected by three different types of edges: 1) code edges, which connect method nodes to statement nodes and among the statement nodes; 2) method call edges, where there is a call dependency between two methods; and 3) coverage edges, where individual methods and their corresponding statements are connected to the tests that cover them. Below, we discuss our graph construction steps. 

\subsubsection{Source Code Graph Construction From Abstract Syntax Trees}

\tool first constructs a graph representation of the code based on the Abstract Syntax Tree (AST) by following prior work~\cite{wang2020detecting, liu2022tell, li2021deeplv, lou2021boosting}, which found that AST is an effective representation of code as inputs for graph neural network. 
Let \( G = (V, E) \) be the graph representation of a program \( p \) under a test method \( t \in T \), where \( T \) is the entire test suite, and \( V \) represents methods and \( E \) represents edges between \( V \). For each \( V \) in program \( p \), we derive an AST, denoted \( V_{ast}\). The nodes in these ASTs typically include constructs such as MethodDeclaration, IfStatement, ReturnStatement, and VariableDeclaration, among others. 
However, not every node in the AST is of equal significance for fault localization. The inclusion of token-level vertices, which represent the finest granularity elements in the AST (e.g., individual operators, literals, and identifiers), introduces overhead without adding considerable information. Thus, we refine the AST by removing these token-level nodes and their corresponding edges, preserving only statement and method declaration-level nodes. This is represented as:
$G_{AST^{\prime}} = G_{AST} - G_{ASTtoken} = (V_{AST^{\prime}},E_{AST^{\prime}})$, where $V_{AST^{\prime}}$ represents set of $\{V_{MethodDeclaration}, V_{ASTstatement} \}$, and  $E_{AST^{\prime}}$ represent edges between $V_{AST^{\prime}}$.
Every method is translated into a root node, \( v_m \in V_M \), which represents the method's \textit{MethodDeclaration} node. Under this root node, statement nodes \( v_s \in V_S \) are structured to capture the code's hierarchy. 

More formally, for each method, we present $V_{MethodDeclaration}$ as the root node, and we represent edge \( e \in E \) such that $ e: V_{ASTstatement} \rightarrow V_{MethodDeclaration} $.
This implies that statement nodes are directly associated with their corresponding method node. Moreover, statement nodes are also interlinked with other statement nodes based on the hierarchical and sequential nature of the AST. This can be represented as:
$ e': v_{s_i} \rightarrow v_{s_j},$ 
where \( v_{s_i} \) and \( v_{s_j} \) are distinct nodes in \( V_S \), and \( e' \) signifies their connection. 

\begin{figure}
    \centering  
    \includegraphics[width=0.8\textwidth]{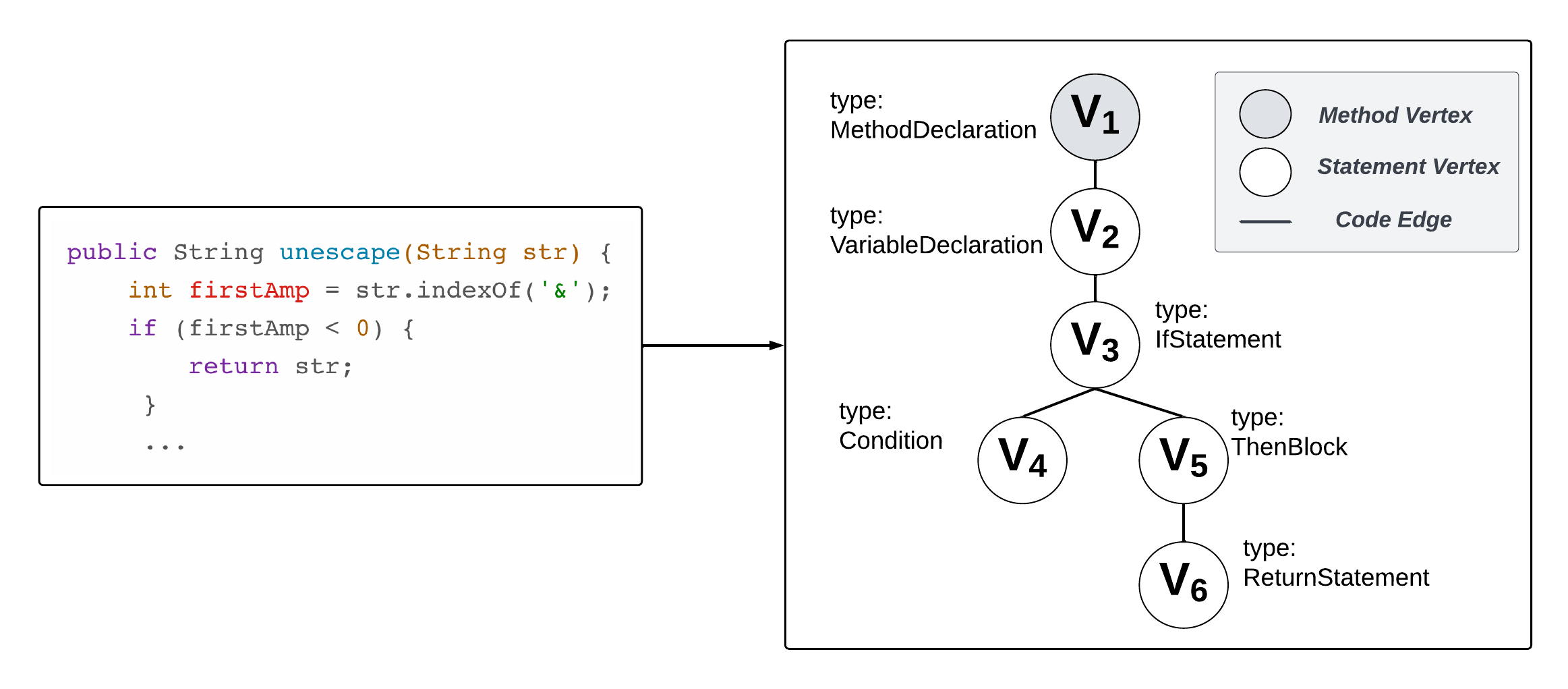}
    \vspace{-0.3cm}
    \caption{Code Representation for the method {\sf unescape()} based on the abstract syntax tree.}
    \label{fig:ast_transform}
    \vspace{-0.3cm}
\end{figure}

The above-mentioned representations ensure that the constructed graph captures the hierarchical structure of the code through the AST while maintaining the semantic structures and dependencies between different parts of the code. Figure \ref{fig:ast_transform} gives an example where a method named \texttt{unescape} is transformed to a code representation following the above technique. This statement-level AST representation, as opposed to individual AST nodes, significantly reduces the number of nodes in the AST, hence reducing the amount of memory consumption. 
Finally, for each $V_{ASTstatement}$, we assign a set of attributes denoted as $attr(V_{ASTstatement})$, which consists of the preserved AST node types. To determine these types, we adopt classifications from Javalang \cite{javalang2023}. This approach recognizes 13 distinct node types, including common constructs like IfStatement and ReturnStatement. Recognizing and categorizing these syntactic constructs provides critical insights, particularly useful in tasks like fault localization. The list of the preserved nodes is publicly available online~\cite{AnonymousSubmission9}.

\subsubsection{Enhancing the Graph with Interprocedural Call Graph Analysis}
\label{sec:callGraph}

To pinpoint the faulty locations, it is essential to understand the flow of execution and the relationships between different parts of the code, as this information is crucial in identifying where faults may originate.
While the AST offers a structured representation of the code, it has limitations in capturing comprehensive interactions between different methods. Consequently, it provides an incomplete view of the program's semantic flow. 
For example, as shown in Figure \ref{fig:full_graph}, ASTs only represent code structures, they cannot identify the flow of execution, such as whether there exists a method call between M2 and M3. To address this limitation, we introduce interprocedural call graph analysis, which gives a clearer picture of the calling dependencies among methods. This analysis extends the structural representation provided by ASTs to show how methods interact.

To better illustrate this analysis, let us define the interprocedural call graph as \( G_{call} = (V_{call}, E_{call}) \). In this graph, \( V_{call} \) represents all the methods, and \( E_{call} \) shows the static method calls. If method \( v_{m1} \) calls method \( v_{m2} \), we represent this with an edge: \( e_{call} = \langle v_{m1},v_{m2} \rangle \), where \( e_{call} \in E_{call} \). This captures how methods interact, both within and between classes. 
We further integrate the interprocedural call graph with the AST-based graph to enhance the code representation. 
The graph is integrated as:
\[G_{integrated} = (V_{AST^{\prime}} \cup V_{call}, E_{AST^{\prime}} \cup E_{call}),\]
which merges the detailed structure from the ASTs with the method calls from the static call graph. As shown in Figure \ref{fig:full_graph}, the integrated graph shows that there is no method call between M2 and M3.
Hence, the interprocedural call graph provides a better understanding of how different methods interact with each other.

\subsubsection{Enhancing Static Graph with Dynamic Code Coverage Information}

Given the integrated graph (\( G_{integrated}\)) from the previous step, we further integrate the dynamic code coverage information: test case \( T \) and the statements (i.e., \( V_S \)) covered by \( T \). 
For a method \( m \) in the faulty program and the test suite \( T \), let \( C[m, t] \) denote the set of statements in \( m \) that are covered by test \( t \), where \( t \in T \). We denote code coverage through a set of edges, \( E_{cov} \), between the code statement nodes \( V_S \) and test nodes \( V_T \) which can be expressed as:
$E_{cov} = \{ \langle v_s, v_t \rangle | s \in C[m, t], t \in T \},$
where \( v_s \) corresponds to the code statement \( s \) within \( V_S \), and \( v_t \) represents the test \( t \) in \( V_T \). 
As an example, as shown in Figure \ref{fig:full_graph}, test node T1 is connected to statement nodes S2, S23, S24, S25, S40, S41 and S42, which shows that T1 covers these statements from methods M1, M2, and M3. 
Similarly, test node T2 is connected to the statement nodes S23, S24, and S25 from method M2.

To provide further context, we associate each test node \( v_t \) with an attribute, \( \text{attr}(v_t) \), indicating its outcome (i.e., pass ($\checkmark$) or fail ($\times$)). Moreover, following the technique from Grace, to scale the graph, we exclude nodes \( v_m \) and \( v_s \) that are not covered by the failing tests. However, as discussed in Section~\ref{motivational_example}, it still does not scale well due to the size of the coverage spectra, i.e., which contains methods unhelpful for fault localization. Hence, we use interprocedural static call graph from Section~\ref{sec:callGraph} to verify the static reachability of method calls \( e_{call} = \langle v_{m1},v_{m2} \rangle \), where \( e_{call} \in E_{call} \). The final unified coverage is defined as, \( G_{integrated} = (V_{AST} \cup V_{call} \cup V_T \cup V_S, E_{AST} \cup E_{call} \cup E_{cov}) \).

\subsection{Enhancing the Dependency-Enhanced Coverage Graph with Additional Graph Attributes} 
\label{sec:attribute}
One major advantage of graph neural networks is their ability to integrate and process node attributes, thereby enriching the model's understanding and improving its learning capabilities \cite{wu2020comprehensive}. By including these additional attributes in the unified coverage graph, we aim to enhance the precision and efficacy in pinpointing faulty statements. We utilize attributes like Test Correlation, which captures textual similarities between method names and failed test cases, and Code Change Information, offering insights into code modifications and their historical context. Below, we provide a detailed description of these attributes.



\phead{Test Correlation.} Test correlation information has been widely recognized in previous studies as a valuable feature for enhancing fault localization~\cite{zhou2012should,lou2021boosting, li2019deepfl}. Test correlation is based on the idea that methods that have a greater textual similarity with the failed test method are more likely to be faulty. We measure the textual similarity between the names of the source method and test method using the Jaccard distance \cite{niwattanakul2013using} by following prior work \cite{li2019deepfl, lou2021boosting}. We split the words based on camel cases as a preprocessing step prior to computing the similarity score. More formally, given method name (\( w_m \)) and failed test method name (\( w_t \)), we define the similarity as:  $Similarity(w_m, w_t) = \frac{length(w_m \cap w_t)}{length(w_t)}  
$. The equation finds the ratio between the overlapping word tokens ($length( w_m \cap w_t )$) and the total number of tokens in the failed test name ($length( w_t )$). When multiple failed tests are present, it greedily selects the test method with the highest similarity score.

\phead{Code Change Information.} Prior studies~\cite{chen2022useful,sohn2017fluccs,wen2019historical, zimmermann2007predicting} found that code changes provide valuable insights into fault proneness. 
Therefore, to improve the effectiveness of fault localization, we incorporate two metrics based on code changes: \textit{Code Churn} and \textit{Method Modification Count (MMC)}. Based on earlier research \cite{zimmermann2007predicting, nagappan2005use}, we define \textit{Code Churn (CHURN)} as the total net change in lines (i.e., both additions and deletions) for a particular method over a given duration. Meanwhile, the \textit{Method Modification Count (MMC)} is determined by the number of modifications made to the method within the same time frame. We calculate each metric using two-time intervals: (1) the entire method history and (2) recent history (six months before the faulty commit, as suggested by Zimmermann et al. \cite{zimmermann2007predicting}), and incorporate them into our graph. To obtain detailed change information, we employ the \textit{git diff} command, capturing the code differences between the faulty commit and its preceding commits. 
To achieve a more detailed analysis, we identify all changed methods from the commits. Using JavaLang~\cite{javalang2023}, we produce an AST for each method on our candidate list. From these ASTs, we extract the beginning and ending line numbers for every method. Subsequently, we verify if any lines within these ranges changed within our selected time intervals. Then, we use this information as method node attributes. 
\
\subsection{Constructing the Graph Neural Network Model}


We deploy a Gated Graph Neural Network (GGNN) to rank potential faulty methods. The input for this model is the \graphname discussed in Section~\ref{sec:graphRepresentation}, represented as \( G(V, E) \), where \( V \) are the nodes and \( E \) are the edges. The nodes, \( V \), comprise the method nodes, statement nodes, and test nodes, which can be expressed as \( V = V_M \cup V_S \cup V_T \). Their interconnections are depicted by the adjacency matrix \( A \). A connection between node \( V_i \) and \( V_j \) is denoted with \( a_{ij} = 1 \). The absence of a connection is marked as \( a_{ij} = 0 \). For the method nodes, each \( v \) in \( V \) carries an attribute \( \gamma_{v} \in \Upsilon \) which includes node type (e.g., \texttt{If Statement}), Test correlation information, Code churn values, Method modification counts. On the other hand, the test vertices contain a test outcome attribute, indicating whether the test passed or failed. 
To ensure stable model performance, we normalize \( A \) with the formula \(\hat{A} = D^{-\frac{1}{2}} A D^{-\frac{1}{2}}\), where \( D \) represents a diagonal matrix. The detailed attribute sequence \( \Upsilon \) along with the adjusted adjacency matrix \( \hat{A} \) are then channeled as primary inputs into the GGNN for the fault detection process.

\phead{Gated Graph Neural Network (GGNN)}. The Gated Graph Neural Network (GGNN) \cite{li2015gated}, an advanced variant of GNNs, incorporates gating mechanisms such as Gated Recurrent Unit (GRU) \cite{dey2017gate} and Long Short-Term Memory (LSTM) \cite{graves2012long} to capture intricate patterns over extended sequences. In \tool, we leverage a GGNN framework to distill essential features from the comprehensive coverage graph in order to localize suspicious statements.

\uhead{\em Embedding Layer}. The embedding layer is designed to transform raw data attributes, represented by the sequence \( \Upsilon \), into a structured and enriched feature matrix, \( \mathcal{X} \), of size \( R^{|V| \times d} \), with \( d \) being the embedding dimension. This transformation aids in offering a better representation of the model's understanding. Test nodes and non-root code nodes are directly transformed into \( d \)-dimensional vectors. For primary code nodes that have more complex attributes, such as the AST node type, a two-step process is adopted. Initially, the AST node type is converted into a \( d-1 \)-dimensional vector. This vector is then concatenated with the test correlation and code change information to complete its representation in the \( d \)-dimensional space. This procedure ensures a uniform \( d \)-dimensional representation for all vertices within the attribute matrix.

\uhead{\em GGNN Iterative Process:} Following a prior work \cite{lou2021boosting}, we apply five iterations in the GGNN layer. This process iteratively refines node representations, enhancing the model's ability to detect patterns and relationships, ultimately aiming for improved results. During the \(t^{th}\) iteration, every node updates its current state by incorporating information from both its neighboring vertices and the outcomes of previous iterations. The gated mechanism in the GGNN is implemented by utilizing the input and forget gates of the LSTM, guiding the propagation of cell states. Specifically, the cell state of node \(v\) in the \(t^{th}\) iteration is represented as \(c^{(t)}_v\), and is initialized as \(c^{(1)}_v = X_v\). 
The propagation of cell states from adjacent vertices in the previous iteration is described by:
\[ a^{(t)}_v = \hat{A}^v:[c^{(t-1)T}_1 ; \ldots ; c^{(t-1)T}_{|V|} ] \]

The forget gates determine which portions of information to exclude from the cell states, while input gates dictate the new information from the current input \(a^{(t)}_v\) to be integrated into the cell states. 
To counteract the vanishing gradient issue, we use the residual connections~\cite{he2016deep} with layer normalization~\cite{ba2016layer} implemented between each pair of sub-layers.

\uhead{\em Inference and Loss Functions in GGNN Analysis}. After processing the graph through the GGNN, our primary goal is to deduce a suspiciousness score for each code statement. The transformation layers, specifically the softmax function, translate the refined node features into these scores.  At a high level, we process the outcomes of the GGNN to generate a suspiciousness score, ranging between 0 and 1, for each candidate method. This score essentially quantifies the likelihood of a method being faulty. The last layer of the GGNN is equipped with a linear transformation that employs the softmax function. Specifically, consider a node, \(v_i\). After the final iteration of GGNN, its output is represented by \(z_i\). This output is transformed linearly to yield a real number \(y'_i\), which is articulated as $ y'_i = Wz_i + b $. 
Here, \(W \in R^{d \times 1}\) is the weight matrix guiding the transformation, and \(b \in R\) is the bias term. Following this, we adopt a listwise strategy to rank the nodes. The outputs for all nodes undergo normalization using the softmax function, resulting in:
\[ p(v_i) = \frac{exp\{y'_i\}}{\sum_{j=1}^n exp\{y'_j\}} \]
This gives \(p(v_i)\) as the probability score that indicates the likelihood of the node \(v_i\) being associated with a fault. When it comes to defining the ranking loss function, our tool, \tool, leans heavily towards the listwise approach. It assesses lists as cohesive entities based on the arrangement of their elements. This approach resonates well with \tool's philosophy of treating elements and their interrelationships as a coherent whole. The corresponding listwise loss function is:
\[ L_{list} = -\sum_{i=1}^n g(v_i) \log(p(v_i)) \]
Wherein, \(g(v_i)\) represents the ground truth label for each method node \(v_i \in V_M\), defined as \(1\) if \(v_i\) is faulty, and \(0\) otherwise.




\section{STUDY DESIGN AND RESULTS}\label{section:result}
In this section, we first describe the study design and setup. Then, we present the motivation, approach, and results of the research questions. 



\phead{Benchmark Dataset.}
To answer the RQs, we conducted the experiment on 675 faults from the Defects4J benchmark (V2.0.0)~\citep{just2014defects4j}. Defects4J provides a controlled environment to reproduce faults collected from projects of various types and sizes. Defects4J has also been widely used in prior automated testing research \cite{lou2021boosting, sohn2017fluccs, chen2022useful, zhang2017boosting}. 
We excluded three projects, JacksonDatabind, JxPath, and Chart, from Defects4J in our study since we encountered many execution errors and were not able to collect the test coverage information.
Table \ref{tab:overview} gives detailed information on the projects and the faults that we use in our study. In total, we conducted our study on 14 projects from Defects4J, which contains 675 unique faults with over 1.4K fault-triggering tests (i.e., failing tests that cover the fault). The sizes of the studied projects range from 2K to 90K lines of code. Note that, since a fault may have multiple fault-triggering tests, there are more fault-triggering tests than faults. 


\begin{table}
\caption{An overview of our studied projects from Defects4J v2.0.0. {\em \#Faults}, {\em LOC}, and {\em \#Tests} show the number of faults, lines of code, and tests in each system. {\em Fault-triggering Tests} shows the number of failing tests that trigger the fault. }
\centering

\scalebox{0.6}{

\setlength{\tabcolsep}{1.5cm}
\begin{tabular}{lrrrr}
    \toprule
    \textbf{Project} & \textbf{\#Faults} & \textbf{LOC} & \textbf{\#Tests} & \textbf{Fault-triggering Tests}\\
    
    \midrule
    Cli          & 39       & 4K     & 94     &  66      \\
    Closure      & 174      & 90K    & 7,911  &  545     \\
    Codec        & 18       & 7K     & 206    &  43      \\
    Collections  & 4        & 65K    & 1,286  &  4       \\
    Compress     & 47       & 9K     & 73     &  72      \\
    Csv          & 16       &  2K    & 54     &  24      \\
    Gson         & 18       & 14K    & 720    &  34      \\
    JacksonCore  & 26       & 22K    & 206    &  53      \\
    JacksonXml   & 6        & 9K     & 138    &  12      \\
    Jsoup        & 93       & 8K     & 139    &  144     \\
    Lang         & 64       & 22K    & 2,291  &  121     \\
    Math         & 106      & 85K    & 4,378  &  176     \\
    Mockito      & 38       & 11K    & 1,379  &  118     \\
    Time         & 26       & 28K    & 4,041  &  74      \\
    \midrule
    \textbf{Total}& 675 & 380K & 24,302 & 1,486 \\
    \bottomrule
\end{tabular}}
\label{tab:overview}

\vspace{-1em}
\end{table}

\phead{Evaluation Metrics. }
According to prior findings, debugging faults at the class level lacks precision for effective location \cite{kochhar2016practitioners}. Alternatively, pinpointing them at the statement level might be overly detailed, omitting important context \cite{parnin2011automated}. Hence, in keeping with prior work \citep{benton2020effectiveness, b2016learning, li2019deepfl, lou2021boosting, vancsics2021call}, we perform our fault localization process at the method level. We apply the following commonly-used metrics for evaluation:

\uhead{{\em Recall at Top-N}}. The Top-N metric measures the number of faults with at least one faulty program element (in this experiment, methods) ranked in the top N. In other words, the Top-N metric identifies the methods that are most relevant to a specific fault among the top-ranked N methods. The result from \tool is a ranked list based on the suspiciousness score. Prior research \cite{parnin2011automated} indicates that developers typically only scrutinize a limited number of top-ranked faulty elements. Therefore, our study focuses on Top-N, where N is set to 1, 3, 5, and 10.

\uhead{\em Mean Average Rank (MAR)}. For each faulty version, the Mean Average Rank (MAR) measures the average position of all faulty methods in a list. For each project, MAR finds the average of these positions across all its faulty versions. The lower the value, the better the ranking. 

\uhead{\em Mean First Rank (MFR)}. This metric determines the position of the first identified fault in the ranked list for every faulty version. The MFR for a project is then derived by taking the average of these positions across all of its faulty versions. A lower MFR means the faulty methods can be identified earlier in the list. 

\phead{Implementation and Environment.}
To gather information about test coverage and calculate the result for baseline techniques, we use Gzoltar \cite{campos2012gzoltar}. Gzoltar is a library for automatic debugging of Java applications. It provides techniques for test suite minimization and fault localization, which can help developers identify and fix software faults more efficiently. Gzoltar can also measure the code coverage of the test suite, which indicates how much of the source code is executed by the tests.  In order to generate static call graphs, we used the Java-Callgraph tool \cite{gousiosg2023javacallgraph}. The tool reads classes from a jar file, walks through their method bodies, and generates a caller-caller relationship table 
to analyze the Java bytecode to retrieve the calling relationship among methods. We then constructed a call graph based on the calling relationship. For generating AST, we parse the source code using the JavaLang toolkit \cite{javalang2023}. 
To train \tool, we use a learning rate of 0.01 and an embedding size of 32 for all projects by following prior studies \cite{li2019deepfl, lou2021boosting}. To prevent under-fitting and over-fitting due to too few or too many epochs, we use 10 epochs according to previous work \cite{li2019deepfl}. To reduce variations across the experiments, we fixed the random seed for all the runs. 
All experiments are conducted on a Linux server with 120G RAM, AMD EPYC 7763 64-Core CPU @ 3.53 GHz, and a 40G NVIDIA A100 GPU. We use PyTorch 2.0.1 for training and validating the GNN model \cite{pytorch}.
The configuration values for our experiment can be found online \cite{AnonymousSubmission9}.

\subsection{RQ1: What is the Effectiveness of DepGraph in Fault Localization?}~\label{sec:rq1}

\phead{Motivation.} In this RQ, we evaluate the localization accuracy of \tool and the impact of leveraging inter-procedural call graph information.
\tool contains two novel components: \graphname and integrating additional node attributes (i.e., code change information) in the graph. Hence, we study the effect of each component separately and the combined effect on fault localization accuracy. The findings may also provide insights into whether a component should be considered in future GNN-based fault localization research. 


\phead{Approach.} 
We compare \tool's fault localization accuracy with three baselines: \textbf{\textit{Ochiai}}~\cite{abreu2006evaluation}, \textbf{\textit{DeepFL}}~\cite{li2019deepfl}, and {\bf \textit{Grace}}~\cite{lou2021boosting} (which we refer to as {\bf \textit{GNN}}). 

\noindent \underline{{\em Ochiai:}} \textit{Ochiai} is a widely used spectrum-based fault localization technique because of its high fault localization accuracy~\cite{abreu2006evaluation, lou2021boosting, li2021fault, cui2020improving, wen2019historical, qian2021agfl}. 
\textit{Ochiai} is defined as: \[Ochiai(a_{ef}, a_{nf}, a_{ep}) = \frac{a_{ef}}{\sqrt{(a_{ef} + a_{nf}) \times (a_{ef} + a_{ep})}}\] where \(a_{ef}\) is the number of failed test cases that execute a statement, \(a_{nf}\) is the number of failed test cases that do not execute the statement, and \(a_{ep}\) is the number of passed test cases that execute the statement. Intuitively, \textit{Ochiai} assigns a higher suspiciousness score to a statement that is executed more frequently by failed test cases and less frequently by passed test cases. The result of \textit{Ochiai} is a value between 0 and 1, with higher values indicating a higher likelihood that a particular statement is faulty. We then aggregate the scores for every statement in a method to get the method-level score.

\noindent \underline{{\em DeepFL:}} \textit{DeepFL}~\cite{li2019deepfl} is a deep-learning-based fault localization technique that integrates spectrum-based and other metrics such as 
code complexity, and textual similarity features to locate faults. It utilizes a Multi-layer Perceptron (MLP) model to analyze these varied feature dimensions. We follow the study~\cite{li2019deepfl} to implement \textit{DeepFL} and include the SBFL scores from 34 techniques, code complexity, and textual similarities as part of the features for the deep learning model.

\noindent \underline{{\em GNN (i.e., Grace):}} \textit{Grace} is one of the first fault localization techniques based on graph neural networks (GNN)~\cite{lou2021boosting} and can be viewed as a direct adoption of GNN. Since \tool is also based on \textit{GNN/Grace}, we study the effectiveness of different components in \tool over \textit{Grace} and refer to \textit{Grace} as \textit{GNN}. \textit{GNN} represents test cases as test nodes in the coverage graph and establishes edges between these test nodes and statement nodes based on test coverage. However, compared to \tool, \textit{GNN’s} graph representation misses call graph info and direct edges between method nodes, which could capture fault propagation. This omission means that \textit{GNN} does not capture potential fault propagation paths between methods, so the candidate method lists produced by \textit{GNN} might include some unrelated methods as discussed in Section \ref{sec:related}. Since \tool can be viewed as an enhanced version of \textit{GNN} (incorporating a \graphnameLower and additional node attributes), our comparison with \textit{GNN} effectively serves as an ablation study.


\noindent \underline{{\em \tool:}} %
In addition to studying the overall fault localization accuracy of \tool, we also construct sub-models to understand the effect of the enhanced graph and node attributes. Specifically, we build two versions of \tool: 1) \toolwocc, which is \textit{GNN + Dependency-Enhanced Coverage Graph} without the code change attributes; and 2) \tool, which incorporates both \graphname and code change information.




\noindent \underline{{\em Evaluating the Models:}} We follow prior studies~\cite{lou2021boosting, li2019deepfl} and evaluate \textit{DeepFL}, \textit{GNN}, and \tool using leave-one-out cross-validation. Given one particular fault, we train the model using all other faults and evaluate the model using that one fault. The process is repeated for all the faults in a project.

\phead{Results.}
\textbf{{\em Across all the studied projects, \toolwocc provides 13\%, 46\%, and 42\% improvement on Top-1, MFR, and MAR, respectively, when compared to GNN.}} Table \ref{tab:rq1} presents the fault localization result of \tool, \toolwocc, and the baseline techniques. Out of 675 faults, \toolwocc identifies 336 faults in Top-1, which is 13\% more than \textit{GNN} (298 faults), 31\% more than \textit{DeepFL} (257), and, 177\% more than \textit{Ochiai} (121 faults). \toolwocc also provides an effective ranked list of potential faulty methods, which can identify 588/675 (87\%) of the faults in Top-10. We see a larger improvement in MFR and MAR when comparing \toolwocc to \textit{GNN}, where the total MFR and MAR are improved by 46\% and 42\%, respectively. The result suggests that the \graphname is effective in promoting faulty methods in the ranked list. Similarly, when comparing \toolwocc to \textit{DeepFL}, there is a significant improvement in both MFR and MAR (65\% and 62\% better, respectively). Among the three techniques, \textit{Ochiai} has the worst MFR and MAR (20.75 and 24.04), which shows that GNN-based techniques are more effective than traditional techniques such as \textit{Ochiai}.

\begin{table} 
    \caption{A comparison of the fault localization techniques. For each project, we show the technique with the best MFR in bold (the lower the better). \toolwocc shows the result after adopting \graphname, and \tool shows the result of incorporating both \graphname and Code Change Information. The number in the parentheses shows the percentage improvement over \textit{GNN (i.e., Grace)}~\cite{lou2021boosting}. The best result is marked in bold.}
    \vspace{-0.3cm}
    \centering
    \label{tab:rq1}
    \scalebox{0.53}{
\setlength{\tabcolsep}{0.6cm}
    \begin{tabular}{l|l|cccccc}
    
    \toprule
        \textbf{Project (\# faults)} & \textbf{Techniques} & \textbf{Top-1} & \textbf{Top-3} & \textbf{Top-5} & \textbf{Top-10} & \textbf{MFR} & \textbf{MAR} \\ \midrule
        \textbf{Cli} (39) & Ochiai & 3 & 5 & 10 & 18 & 15.711 & 18.272 \\ 
        \textbf{} & DeepFL & 11 & 21 & 24 & 28 & 8.991 & 10.681 \\ 
        \textbf{} & GNN & 14 & 24 & 26 & 30 & 7.861 & 9.903 \\ 
        \textbf{} & \toolwocctable & 15 (7\%) & 22 (-8\%) & 26 (0\%) & 31 (3\%) & 5.973 (24\%) & 7.118 (28\%) \\ 
        \textbf{} & \tooltable & \textbf{17 (21\%)} & 24 (0\%) & \textbf{27 (4\%)} & \textbf{34 (10\%)} & \textbf{5.105 (30\%)} & \textbf{6.223 (32\%)} \\ 
        \midrule
        \textbf{Closure} (174) & Ochiai & 20 & 39 & 70 & 72 & 98.652 & 110.348 \\ 
        \textbf{} & DeepFL & 46 & 61 & 92 & 99 & 29.388 & 35.333 \\ 
        \textbf{} & GNN & 51 & 78 & 102 & 121 & 12.854 & 14.814 \\ 
        \textbf{} & \toolwocctable & 58 (14\%) & 97 (24\%) & 123 (21\%) & \textbf{148 (22\%)} & 4.844 (62\%) & 7.911 (47\%) \\ 
        \textbf{} & \tooltable & \textbf{60 (18\%)} & \textbf{99 (27\%)} & \textbf{126 (24\%)} & \textbf{148 (22\%)} & \textbf{4.542 (65\%)} & \textbf{7.306 (51\%)} \\ 
        \midrule
        \textbf{Codec} (18) & Ochiai & 3 & 12 & \textbf{17} & 17 & 2.701 & 3.461 \\ 
        \textbf{} & DeepFL & 5 & 10 & 12 & 16 & 2.742 & 4.803 \\ 
        \textbf{} & GNN & 6 & 11 & 13 & 17 & 2.536 & 4.015 \\ 
        \textbf{} & \toolwocctable & \textbf{7 (17\%)} & \textbf{12 (9\%)} & 14 (8\%) & 16 (-6\%) & \textbf{2.412 (5\%)} & \textbf{3.265 (19\%)} \\ 
        \textbf{} & \tooltable & \textbf{7 (17\%)} & 10 (-9\%) & 14 (8\%) & 16 (-6\%) & 3.111 (-23\%) & 4.327 (-8\%) \\ 
        \midrule
        \textbf{Collections} (4) & Ochiai & 1 & 1 & 2 & 2 & 3.871 & 3.431 \\ 
        \textbf{} & DeepFL & 1 & 1 & 2 & 2 & 1.512 & 1.519 \\ 
        \textbf{} & GNN & 1 & 1 & 2 & 2 & 1.511 & 1.511 \\ 
        \textbf{} & \toolwocctable & \textbf{1 (0\%)} & 1 (0\%) & \textbf{2 (0\%)} & 2 (0\%) & 1.511 (0\%) & 1.511 (0\%) \\ 
        \textbf{} & \tooltable & \textbf{1 (0\%)} & \textbf{2 (100\%)} & \textbf{2 (0\%)} & 2 (0\%) & \textbf{1.445 (4\%)} & \textbf{1.445 (4\%)} \\ 
        \midrule
        \textbf{Compress} (47) & Ochiai & 5 & 12 & 17 & 29 & 20.106 & 23.275 \\ 
        \textbf{} & DeepFL & 22 & 27 & 31 & 38 & 9.573 & 12.955 \\ 
        \textbf{} & GNN & 23 & 29 & 34 & 42 & 5.383 & 6.987 \\ 
        \textbf{} & \toolwocctable & 24 (4\%) & 32 (10\%) & \textbf{36 (6\%)} & 42 (0\%) & 4.384 (19\%) & 5.209 (25\%) \\ 
        \textbf{} & \tooltable & \textbf{25 (9\%)} & \textbf{33 (14\%)} & \textbf{36 (6\%)} & \textbf{45 (7\%)} & \textbf{3.361 (38\%)} & \textbf{4.245 (39\%)} \\ 
        \midrule
        \textbf{Csv} (16) & Ochiai & 3 & 8 & 10 & 12 & 5.625 & 5.782 \\ 
        \textbf{} & DeepFL & 7 & 8 & 9 & 11 & 5.623 & 5.971 \\ 
        \textbf{} & GNN & 6 & 8 & 10 & 12 & 5.438 & 5.938 \\ 
        \textbf{} & \toolwocctable & \textbf{8 (33\%)} & \textbf{9 (13\%)} & 10 (0\%) & \textbf{13 (8\%)} & 5.362 (1\%) & 5.581 (6\%) \\ 
        \textbf{} & \tooltable & \textbf{8 (33\%)} & \textbf{9 (13\%)} & \textbf{12 (20\%)} & \textbf{13 (8\%)} & \textbf{4.813 (12\%)} & \textbf{5.001 (16\%)} \\ 
        \midrule
        \textbf{Gson} (18) & Ochiai & 4 & 9 & 9 & 12 & 9.177 & 10.183 \\ 
        \textbf{} & DeepFL & 8 & 11 & 12 & 12 & 8.873 & 9.324 \\ 
        \textbf{} & GNN & 11 & 13 & 14 & 15 & 6.471 & 6.755 \\ 
        \textbf{} & \toolwocctable & 12 (9\%) & 14 (8\%) & 15 (7\%) & 15 (0\%) & 2.177 (66\%) & 2.471 (63\%) \\ 
        \textbf{} & \tooltable & \textbf{14 (27\%)} & \textbf{15 (15\%)} & \textbf{16 (14\%)} & \textbf{16 (7\%)} & \textbf{1.353 (79\%)} & \textbf{1.765 (74\%)} \\ 
        \midrule
        \textbf{JacksonCore} (26) & Ochiai & 6 & 11 & 13 & 14 & 9.789 & 16.754 \\ 
        \textbf{} & DeepFL & 5 & 5 & 9 & 10 & 8.671 & 9.711 \\ 
        \textbf{} & GNN & 9 & 13 & 14 & 15 & 6.471 & 6.755 \\ 
        \textbf{} & \toolwocctable & 9 (0\%) & 14 (8\%) & \textbf{15 (7\%)} & \textbf{17 (13\%)} & 3.474 (29\%) & 4.509 (28\%) \\ 
        \textbf{} & \tooltable & \textbf{12 (33\%)} & \textbf{15 (15\%)} & \textbf{15 (7\%)} & 16 (7\%) & \textbf{3.052 (38\%)} & \textbf{4.015 (36\%)} \\ 
        \midrule
        \textbf{JacksonXml} (6) & Ochiai & 0 & 0 & 0 & 0 & 59.2 & 59.2 \\ 
        \textbf{} & DeepFL & 3 & 3 & 4 & 5 & 3.513 & 4.245 \\ 
        \textbf{} & GNN & 3 & 3 & 4 & 5 & 2.401 & 2.401 \\ 
        \textbf{} & \toolwocctable & \textbf{4 (33\%)} & \textbf{5 (67\%)} & \textbf{5 (25\%)} & \textbf{5 (0\%)} & 0.411 (83\%) & 0.411 (83\%) \\ 
        \textbf{} & \tooltable & \textbf{4 (33\%)} & \textbf{5 (67\%)} & \textbf{5 (25\%)} & \textbf{5 (0\%)} & \textbf{0.409 (83\%)} & \textbf{0.409 (83\%)} \\ 
        \midrule
        \textbf{Jsoup} (93) & Ochiai & 15 & 40 & 48 & 57 & 14.944 & 20.209 \\ 
        \textbf{} & DeepFL & 33 & 39 & 46 & 49 & 10.23 & 11.444 \\ 
        \textbf{} & GNN & 40 & 64 & 72 & 77 & 8.223 & 9.669 \\ 
        \textbf{} & \toolwocctable & 50 (25\%) & 70 (9\%) & 77 (7\%) & 82 (6\%) & 4.022 (51\%) & 6.815 (30\%) \\ 
        \textbf{} & \tooltable & \textbf{53 (33\%)} & \textbf{73 (14\%)} & \textbf{78 (8\%)} & \textbf{83 (8\%)} & \textbf{3.023 (63\%)} & \textbf{4.6174 (52\%)} \\ 
        \midrule
        \textbf{Lang} (64) & Ochiai & 25 & 45 & 51 & 59 & 4.68 & 5.15 \\ 
        \textbf{} & DeepFL & 42 & 53 & 55 & 57 & 2.833 & 3.08 \\ 
        \textbf{} & GNN & 43 & 53 & 57 & 58 & 2.113 & 2.462 \\ 
        \textbf{} & \toolwocctable & 45 (5\%) & \textbf{55 (4\%)} & 58 (2\%) & \textbf{61 (5\%)} & 1.564 (26\%) & 1.902 (23\%) \\ 
        \textbf{} & \tooltable & \textbf{48 (12\%)} & \textbf{55 (4\%)} & \textbf{60 (5\%)} & \textbf{61 (5\%)} & \textbf{1.153 (45\%)} & \textbf{1.481 (40\%)} \\ 
        \midrule
        \textbf{Math} (106) & Ochiai & 23 & 52 & 62 & 82 & 9.73 & 11.72 \\ 
        \textbf{} & DeepFL & 52 & 81 & 90 & 95 & 3.95 & 4.911 \\ 
        \textbf{} & GNN & 64 & 79 & 92 & 97 & 2.355 & 3.082 \\ 
        \textbf{} & \toolwocctable & 67 (5\%) & 90 (14\%) & 96 (4\%) & 100 (3\%) & 1.185 (50\%) & 1.528 (50\%) \\ 
        \textbf{} & \tooltable & \textbf{72 (13\%)} & \textbf{92 (16\%)} & \textbf{97 (5\%)} & \textbf{102 (5\%)} & \textbf{1.115 (53\%)} & \textbf{1.454 (53\%)} \\ 
        \midrule
        \textbf{Mockito} (38) & Ochiai & 7 & 14 & 18 & 23 & 20.22 & 24.77 \\ 
        \textbf{} & DeepFL & 10 & 18 & 23 & 26 & 13.541 & 17.001 \\ 
        \textbf{} & GNN & 16 & 24 & 26 & 29 & 9.611 & 13.621 \\ 
        \textbf{} & \toolwocctable & 20 (25\%) & 28 (17\%) & \textbf{34 (31\%)} & \textbf{34 (17\%)} & 2.361 (75\%) & 3.307 (76\%) \\ 
        \textbf{} & \tooltable & \textbf{21 (31\%)} & \textbf{29 (21\%)} & 32 (23\%) & \textbf{34 (17\%)} & \textbf{2.194 (77\%)} & \textbf{2.998 (78\%)} \\ 
        \midrule
        \textbf{Time} (26) & Ochiai & 6 & 12 & 13 & 16 & 16.14 & 18.98 \\ 
        \textbf{} & DeepFL & 12 & 15 & 18 & 20 & 12.722 & 13.754 \\ 
        \textbf{} & GNN & 11 & 16 & 20 & 21 & 7.842 & 8.448 \\ 
        \textbf{} & \toolwocctable & 16 (45\%) & 19 (19\%) & 20 (0\%) & \textbf{22 (5\%)} & 3.321 (58\%) & 4.465 (47\%) \\ 
        \textbf{} & \tooltable & \textbf{17 (55\%)} & \textbf{20 (25\%)} & \textbf{21 (5\%)} & \textbf{22 (5\%)} & \textbf{3.044 (61\%)} & \textbf{4.371 (48\%)} \\ 
        \midrule
        \textbf{Total} (675) & Ochiai & 121 & 260 & 340 & 413 & 20.753 & 24.038 \\ 
        \textbf{} & DeepFL & 257 & 353 & 427 & 468 & 8.726 & 10.338 \\ 
        \textbf{} & GNN & 298 & 416 & 486 & 541 & 5.678 & 6.851 \\ 
        \textbf{} & \toolwocctable & 336 (13\%) & 470 (13\%) & 534 (10\%) & 588 (9\%) & 3.049 (46\%) & 3.957 (42\%) \\ 
        \textbf{} & \tooltable & \textbf{359 (20\%)} & \textbf{481 (16\%)} & \textbf{541 (11\%)} & \textbf{597 (10\%)} & \textbf{2.562 (55\%)} & \textbf{3.272 (52\%)} \\ 
        \bottomrule
    \end{tabular}
    }
\end{table}

\noindent\textbf{\textit{Across all the studied projects, further adding the code change information to \tool improves the overall Top-1, MFR, and MAR by 7\%, 16\%, and 17\%, respectively, compared to \toolwocc.}} Although the relative improvement is less for Top-3. Top-5, and Top-10 (2.3\%, 1.3\%, and 1.5\%, respectively), code change information is more effective at improving Top-1. \tool can identify 23 more faults in Top-1, representing a 7\% improvement. Having faulty methods ranked earlier in the list is essential for developers’ adoption of fault localization techniques~\cite{parnin2011automated}, which further shows the importance of adding code change information. When looking at individual projects, adding code change information, in general, also gives the best results in all the studied projects. One exception is Codec, where we see a decrease in Top-3, Top-10, MFR, and MAR after adding code change information. The reason may be that the number of faults in Codec is relatively small (18 faults), so missing one fault causes larger fluctuations in the localization result. Our finding shows that code representation and the information in the graph play a significant role in improving the fault localization result. Future studies should adopt enhanced graph representation of the code, and consider combining SBFL techniques with other valuable information that can be mined from the software development history.

\rqboxc{Adding the \graphname to \textit{GNN} improves Top-1 by 20\% and MFR and MAR by more than 50\%. Adding the code change information further improves Top-1 by 7\%. Our findings emphasize improving graph representation and combining SBFL techniques with other information that can be mined from the software development history.}




\subsection{RQ2: How Much Computing Resource Can Be Reduced By Adopting the Dependency-Enhanced Coverage Graph?} 

\phead{Motivation.} Due to the complexity of graph data, training and using graph neural networks (GNNs) require a significant amount of memory (e.g., need to store the entire graph in GPU memory) and computation resources \cite{zhou2022gnnear}. 
In the case of source code, the graph size grows exponentially due to having various control flows, which can make a graph extremely large for real-world software, making GNNs difficult and slow to train. 
Prior works \cite{qian2021agfl, qian2023gnet4fl, nguyen2022ffl, lou2021boosting} in graph-based fault localization only utilize AST trees to represent the code. However, as we discussed in Section~\ref{motivational_example}, representing the code using only the AST trees would miss the caller-callee information among the method nodes, potentially adding unnecessary edges among the nodes. 
In this RQ, we study how much resources can be saved when training the model using the \graphnameLower.

\phead{Approach.} 
To answer this RQ, we evaluated several metrics for the generated graph of each project: number of nodes and edges, training/inference time, and GPU memory usage. 
Our comparison centers on two distinct phases of model configuration. The `before' phase utilizes a graph representation similar to the \textit{Grace (e.g., GNN)} model, which notably does not include inter-procedural call graph information, and serves as the baseline in our analysis. In contrast, the `after' phase mirrors the approach employed in \tool, where we integrate inter-procedural call graph information into the graph representation (i.e., the \graphname).
We measure the response time by comparing the time that we start to train the model and the time that the training is done using both the baseline and \graphnameLower representations. For GPU memory, we examine and report the memory usage when it is stabilized (e.g., once the graph is loaded in the GPU memory and the model training starts).



\phead{Result.}
\noindent{\textit{\textbf{Adopting the \graphnameLower not only improves the localization results as shown in RQ1, but it can also reduce the number of nodes and edges by 69.32\% and 74.93\%, respectively, and the GPU memory usage by 44\%. }}}
Table \ref{tab:combined_comparison_graphsize} shows the graph size reduction before and after adopting the \graphnameLower. 
For all of the projects, the number of nodes and edges decreased significantly (69.32\% and 74.93\% across all projects). We find that such reduction can be even greater for larger projects. For instance, the number of nodes in the largest project Closure is around 728.8K when only using the AST information, but reduces significantly to around 137K after adopting \graphnameLower (81\% reduction). There were also around 100 million edges initially and they were reduced to around 10.5 million (89\% reduction). A similar trend can be observed in other large projects, such as Jsoup, where the node count is reduced from around 131K to 61K (53.31\% reduction), and Math, where the reduction is even greater (74.10\%). 
For smaller projects, the reduction in nodes and edges may be less (e.g., Codec has 25.46\% and 30\% reduction in nodes and edges), but overall, adopting \graphnameLower can significantly reduce graph sizes and improve localization. 

\begin{table}
    \centering
    \caption{Comparisons of graph sizes, training/inference time, and GPU memory usage during training before and after adopting the \graphnameLower. The numbers in the parentheses show the percentage improvement. {\em Total} shows the aggregated results from all the studied projects. }
    \vspace{-0.3cm}
    \label{tab:combined_comparison_graphsize}
    \scalebox{0.7}{
    \setlength{\tabcolsep}{0.12cm}
    \begin{tabular}{l|rr|rr|rr|rr|rr}
    \toprule
    \multirow{2}{*}{\textbf{Project}} & \multicolumn{2}{c|}{\textbf{\#Nodes}} & \multicolumn{2}{c|}{\textbf{\#Edges}} & \multicolumn{2}{c|}{\textbf{Training(s)}} & \multicolumn{2}{c|}{\textbf{Inference(s)}} & \multicolumn{2}{c}{\textbf{GPU Memory Usage(GB)}} \\
    & Before & After & Before & After & Before & After & Before & After & Before & After \\
    \midrule
    Cli & 17.8K & 12.1K (32.31\%) & 782.7K & 229.6K (70.67\%) & 2.2K & 491 (79\%) & 118 & 22 (81\%) & 3.5 & 1.9 (44\%) \\
    Closure & 728.8K & 137.3K (81.16\%) & 100.2M & 10.5M (89.48\%) & 579.6K & 102K (82\%) & 5.2K & 556 (89\%) & 35.6 & 23.9 (33\%) \\
    Codec & 3.4K & 2.5K (25.46\%) & 46.4K & 32.5K (29.99\%) & 223 & 40 (82\%) & 14 & 6 (57\%) & 1 & 0.8 (20\%) \\
    Compress & 30.2K & 14.9K (50.65\%) & 876.1K & 244.8K (72.06\%) & 2.1K & 703 (67\%) & 53 & 33 (38\%) & 4.5 & 1.9 (57\%) \\
    Csv & 5.8K & 3.5K (39.93\%) & 215.0K & 85.2K (60.40\%) & 383 & 172 (55\%) & 27 & 12 (56\%) & 2.7 & 0.9 (66\%) \\
    Gson & 19.6K & 12.6K (35.71\%) & 2.4M & 775.9K (68.23\%) & 2.9K & 675 (77\%) & 162 & 38 (77\%) & 7.4 & 2.9 (60\%) \\
    JacksonCore & 14.7K & 7.1K (51.65\%) & 1.4M & 216.7K (84.22\%) & 2.2K & 429 (81\%) & 173 & 62 (64\%) & 5.7 & 2.1 (81\%) \\
    JacksonXml & 2.9K & 1.0K (64.22\%) & 163.0K & 23.4K (85.64\%) & 79 & 19 (75\%) & 12 & 5 (57\%) & 2.3 & 1.3 (44\%) \\
    Jsoup & 131.4K & 61.3K (53.31\%) & 22.3M & 4.4M (80.29\%) & 137K & 21K (84\%) & 1.3K & 232 (82\%) & 23 & 11.1 (51\%) \\
    Lang & 18.1K & 9.3K (48.75\%) & 435.6K & 142.2K (67.35\%) & 1.7K & 758 (57\%) & 52 & 29 (44\%) & 1.8 & 1.2 (34\%) \\
    Math & 140.9K & 36.5K (74.10\%) & 3.3M & 1.1M (67.26\%) & 25K & 7.3K (71\%) & 234 & 95 (59\%) & 18 & 10.3 (43\%) \\
    Mockito & 63.3K & 28.0K (55.69\%) & 10.7M & 521.2K (95.14\%) & 31.4K & 1K (97\%) & 1K & 37 (97\%) & 19.3 & 5.2 (73\%) \\
    Time & 99.1K & 65.3K (34.15\%) & 4.9M & 4.2M (14.89\%) & 10.4K & 5K (52\%) & 329 & 220 (33\%) & 18.5 & 15.7 (15\%) \\
    \midrule
    {\bf Total} & 1.27M & 391.4K (69.32\%) & 47.54M & 11.91M (74.93\%) & 796K & 140K (82\%) & 9K & 1.3K (85\%) & 143.7 & 80 (44\%) \\
    \bottomrule
    \end{tabular}
    }\vspace{-0.3cm}
\end{table}

The reduction in graph size also contributes to a significant improvement in GPU memory usage. We see an improvement ranging from 15\% to 81\% in GPU memory usage across all the studied projects. When summing up the usage across all the studied projects, adopting \graphnameLower reduces the memory usage from 143.7GB to 80GB (44\% improvement). For larger projects such as Closure, we can reduce more than 10GB of GPU memory (from 35.6GB to 23.9GB) in the training process. Reducing the GPU memory usage is significant for the practicability and adoption of all GNN-based techniques since a high-end GPU such as NVIDIA A100 only has 40GB or 80GB of memory. Hence, reducing memory usage can make GNN-based techniques easier to adopt on larger projects.


\noindent\textbf{{\em The model training and inference time are reduced by 82\% and 85\%, respectively, after adopting the \graphnameLower.}} The training time for the GNN model depends on the size of the project. However, in our experiment, the training time is significant, with a total training time of over 9 days for all the projects (796K seconds) on an NVIDIA A100 GPU before adopting the \graphnameLower. For larger projects like Closure, the training time took almost one week. Such a long training time may hinder the adoption of GNN-based techniques in general and is not limited only to fault localization. 
After adopting the \graphnameLower, we noticed a significant reduction in the training time, where it took only one day to train a model for Closure (82\% reduction). 
Across all the studied projects, we find that the total training time is reduced by 82\%, where the improvement is at least 50\% or more for individual projects. 
Although inference is faster than training, the total inference time still takes 2.5 hours (9K seconds). In particular, the inference time is 1.4 hours for Closure. After adopting the \graphnameLower, the total inference time is reduced by 85\%, from 2.5 hours to 20 minutes (1.3K).

Our findings highlight the challenges in adopting GNN-based techniques (not just fault localization) in practice while also showing potential future directions.  
Graph neural networks (GNNs) typically compute an entire adjacency matrix along with the embedding for all nodes. This process can be notably resource-intensive, both in terms of memory and computational time, as found in our study and also highlighted in prior work~\cite{ma2022graph}. In this paper, we introduced a more compact and more accurate graph representation that substantially reduces the overall size of the graph. This reduction, in turn, has led to a great reduction decrease in both training and inference time, and memory usage. As demonstrated in RQ1, our approach does not compromise fault localization accuracy yet it actually improves the accuracy. 
Future studies should consider improving the practical aspects of GNN-based fault localization by further improving the graph representation and perhaps incorporating other graph pruning techniques.



\rqboxc{Adopting the \graphname helps significantly reduce the graph size by 70\% and GPU memory usage by 44\%. We also find that the total training time is reduced from 9 days to 1 day. Our findings highlight the computation issue with using GNN techniques for software engineering tasks in general and provide potential future research directions. }

\subsection{RQ3: Does DepGraph Locate Different Sets of Faults Compared to GNN?} 

\phead{Motivation.} In this RQ, 
we further evaluate what kind of faults \tool is able to detect. Specifically, whether \tool can locate an additional set of faults, or does it detects new faults and misses some previously-located faults. 
Understanding such changes in the located faults may help us understand why our approach works and fails for certain faults and the effect of \graphnameLower and code change information. The findings may give us valuable insight into improving GNN-based FL techniques in the future. 

\phead{Approach.} We conduct two experiments on the two representative code representations (\graphname and the representation used in GNN (i.e., Grace)~\cite{lou2021boosting}), and the code change information we added: (1) we analyze the degree of overlap between the faults located by different code representation at Top-N, and additionally show whether there are overlaps in the located faults; and (2) from these overlapping and non-overlapping faults, we perform an empirical study to analyze what kind of faults each code representation is capable of locating at Top-N. For a given fault, we extract their faulty statements and analyze their (a) AST node type and (b) fix information, to identify the relationship between code presentation and fault location. 

\begin{figure}
    \centering 
    \includegraphics[width=0.8\columnwidth]{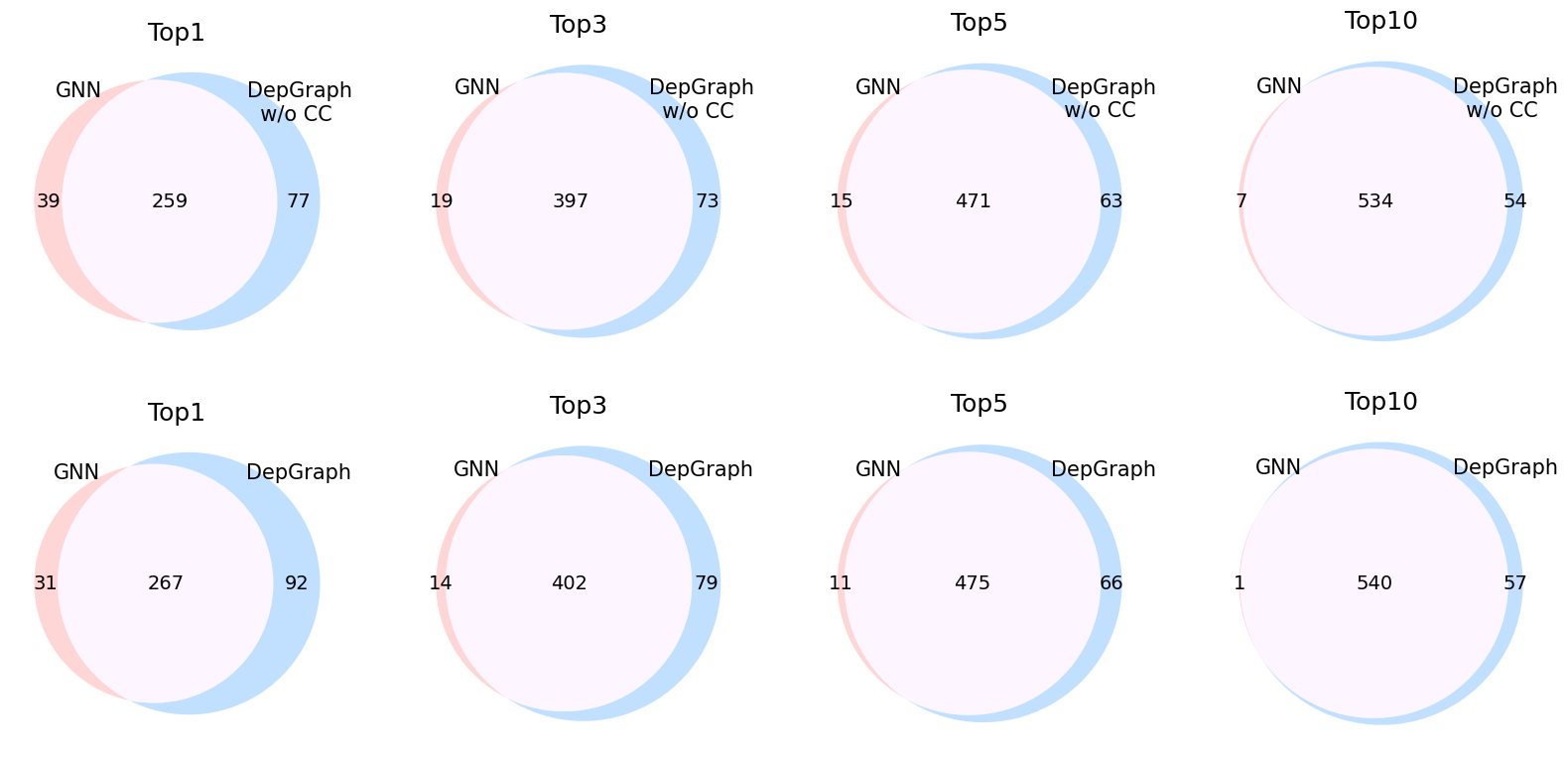} 
    \caption{Overlaps between the faults that \tool and \textit{GNN (i.e., Grace)} locate in Top-1, 3, 5 and 10. The overlapping regions contain a number of faults that have the same ranking in both of the techniques. The non-overlapped regions contain faults that were uniquely located by each technique. We consider both \tool, and, \textit{DepGraph w/o CC (Code Change)}.}
    \label{fig:unmatch}
\end{figure}

\phead{Results.} \textit{\textbf{Adopting \graphnameLower helps locate 
8.8\% to 13.6\% additional faults compared to \textit{GNN}. By further adding code change information, we can locate 10\% to 26\% additional faults.}} In our analysis, we compare the faults that can be located using different graph representations: \textit{GNN}, \toolwocc (with call dependency), and \tool. As shown in Figure~\ref{fig:unmatch}, we observe that \toolwocc is able to locate 77 additional new faults at Top-1 that could not be located by \textit{GNN} (14.7\% new faults). 
At Top-3 we see that \toolwocc can still locate 73 additional faults compared to \textit{GNN} (13.6 \% new faults), 63 additional faults (9.6\%) at top-5, and 54 additional faults (8.8\%). These results show that as we increase Top-N, \toolwocc can locate almost all the faults that \textit{GNN} can locate, but at the same time, \toolwocc can locate many additional faults.  
Similarly, when adding code change information, \tool is able to locate more faults, and at the same time, miss fewer faults that were previously detected when using \textit{GNN}. 
Moreover, at Top-10 we find that \tool can locate almost all the faults that \textit{GNN} located, except for one. The findings show the robustness of adding additional information to the graph. 
Adding \graphnameLower provides mainly benefits in locating additional faults and does not miss the previously-located faults.

\noindent\textit{\textbf{Compared to \textit{GNN}, the additional faults that \tool locates are related to method-to-method relationship. The findings show the importance and effectiveness of our \graphnameLower. }} While our \tool can locate more faults compared to \textit{GNN}, there are still a few faults that our technique cannot locate (and vice versa). We manually analyze this discrepancy to help us understand whether our techniques are superior in locating certain types of faults over \textit{GNN}. Upon analyzing faults in Lang, we find that \tool may be more effective in locating faults that are associated with the loop structures like \textit{``for''} or \textit{``while''} loops or control statements. For instance, in Lang, \tool found eight unique faults, while \textit{GNN} found only two. When we checked those two faults located by \textit{GNN}, we observed that neither fault had loop or control type statements. Moreover, out of the eight faults that were located by \tool, four were related to fixing \textit{``for''} or \textit{``while''} loop. 
Additionally, \tool exhibits peculiarity in locating faults associated with method-call relationships. For example, in JacksonCore, \tool located two additional faults compared to \textit{GNN}. One of the two faults was related to an incorrect method-to-method call between the two classes. Hence, the analysis shows that enhancing graph representation with call graph improves the ability of our \tool to locate certain types of faults that \textit{GNN} cannot locate. 
 
\rqboxc{Adopting \graphnameLower and code change information helps locate 10\% to 26\% additional faults compared to \textit{GNN}. We also find that \tool is able to detect faults that are related to method-to-method relationships, loop structures, and method-call relationships, which \textit{GNN} cannot locate.}

\subsection{RQ4: What is the Cross-Project Fault Localization Accuracy?} 

\begin{table} 
    \caption{A comparison of the cross-project fault localization accuracy. In the last row, the number in the parentheses shows the percentage improvement over {\it GNN} (i.e., {\it Grace})~\cite{lou2021boosting}. The best result is marked in bold.}
    \vspace{-0.3cm}
    \centering
    \label{tab:rq4}
    \scalebox{0.6}{
\setlength{\tabcolsep}{0.28cm}
    \begin{tabular}{l|l|cccccc}
    
    \toprule
        \textbf{Project (\# faults)} & \textbf{Techniques} & \textbf{Top-1} & \textbf{Top-3} & \textbf{Top-5} & \textbf{Top-10} & \textbf{MFR} & \textbf{MAR} \\ \midrule
        \textbf{Cli} (39) & GNN & 9.25 & 14.917 & 17.5 & 21.667 & 15.156 & 17.694 \\ 
        \textbf{} & \toolwocctable & 9.846 & 17.462 & 21 & 26.846 & 8.13 & 9.511 \\ 
        \textbf{} & \tooltable & \textbf{13.083} & \textbf{19.583} & \textbf{22.417} & \textbf{28.417} & \textbf{6.702} & \textbf{8.011} \\ 
        \midrule
        \textbf{Closure} (174) & GNN & 32.456 & 63.986 & 78.492 & 94.432 & 24.611 & 29.011 \\ 
        \textbf{} & \toolwocctable & 38.231 & 70.231 & 92.538 & 115.923 & 17.55 & 22.039 \\ 
        \textbf{} & \tooltable & \textbf{41.545} & \textbf{73.111} & \textbf{97.736} & \textbf{116.672} & \textbf{14.753} & \textbf{21.555} \\ 
        \midrule
        \textbf{Codec} (18) & GNN & 5.333 & 10.5 & 12.083 & 14.917 & 4.093 & 6.21 \\ 
        \textbf{} & \toolwocctable & 5.923 & \textbf{11.385} & \textbf{13} & 15.462 & 3.208 & \textbf{4.489} \\ 
        \textbf{} & \tooltable & \textbf{6.667} & 10.667 & 12.583 & \textbf{16.083} & \textbf{2.99} & 5.549 \\ 
        \midrule
        \textbf{Collections} (4) & GNN & 1.25 & 1.5 & 1.833 & 1.833 & 2.958 & 2.958 \\ 
        \textbf{} & \toolwocctable & 1.154 & 1.538 & 1.923 & 2 & 1.192 & 1.192 \\ 
        \textbf{} & \tooltable & \textbf{1.667} & \textbf{2} & \textbf{2} & \textbf{2} & \textbf{0.25} & \textbf{0.25} \\ 
        \midrule
        \textbf{Compress} (47) & GNN & 14.833 & 23.833 & 28 & 33.417 & 11.406 & 13.686 \\ 
        \textbf{} & \toolwocctable & 17.846 & 27.846 & 32.077 & 37.769 & 6.494 & 7.73 \\ 
        \textbf{} & \tooltable & \textbf{21.417} & \textbf{31.833} & \textbf{36.417} & \textbf{41.333} & \textbf{4.619} & \textbf{5.696} \\ 
        \midrule
        \textbf{Csv} (16) & GNN & 5.583 & 7.75 & 9.167 & 10.75 & 10.578 & 11.083 \\ 
        \textbf{} & \toolwocctable & 6 & \textbf{8.615} & \textbf{10.769} & \textbf{12.923} & \textbf{5.947} & \textbf{6.106} \\ 
        \textbf{} & \tooltable & \textbf{6.75} & 8.417 & 10.5 & 12.5 & 7.203 & 7.451 \\ 
        \midrule
        \textbf{Gson} (18) & GNN & 7.417 & 10.167 & 11.75 & 13.167 & 13.784 & 14.23 \\ 
        \textbf{} & \toolwocctable & \textbf{9.231} & 11.615 & 12.692 & 14.308 & \textbf{4.195} & \textbf{4.509} \\ 
        \textbf{} & \tooltable & 7.917 & \textbf{11.667} & \textbf{13.5} & \textbf{15} & 6.475 & 6.739 \\ 
        \midrule
        \textbf{JacksonCore} (26) & GNN & 0.917 & 3.167 & 3.667 & 5.417 & 41.851 & 47.374 \\ 
        \textbf{} & \toolwocctable & 7.154 & 12.462 & 13.462 & \textbf{15.385} & \textbf{4.822} & \textbf{6.171} \\ 
        \textbf{} & \tooltable & \textbf{8.667} & \textbf{12.667} & \textbf{13.833} & 14.75 & 8.246 & 9.823 \\ 
        \midrule
        \textbf{JacksonXml} (6) & GNN & 0.75 & 1.667 & 1.917 & 3 & 25.733 & 25.733 \\ 
        \textbf{} & \toolwocctable & 1.538 & 2.615 & 3.077 & 4.154 & 4.215 & 4.215 \\ 
        \textbf{} & \tooltable & \textbf{2} & \textbf{3.333} & \textbf{4.333} & \textbf{4.833} & \textbf{2.15} & \textbf{2.15} \\ 
        \midrule
        \textbf{Jsoup} (93) & GNN & 25.667 & 42.833 & 53.417 & 61.667 & 29.87 & 36.572 \\ 
        \textbf{} & \toolwocctable & 36.231 & 50.154 & 61.462 & 69.846 & 11.028 & 13.879 \\ 
        \textbf{} & \tooltable & \textbf{43.25} & \textbf{60.583} & \textbf{68.917} & \textbf{77.5} & \textbf{4.766} & \textbf{7.503} \\ 
        \midrule
        \textbf{Lang} (64) & GNN & 34.583 & 48.25 & 52.583 & 57.333 & 3.035 & 3.584 \\ 
        \textbf{} & \toolwocctable & 37.462 & 50.077 & 54.538 & 60 & 1.935 & 2.241 \\ 
        \textbf{} & \tooltable & \textbf{42.667} & \textbf{53.583} & \textbf{57.75} & \textbf{60.75} & \textbf{1.507} & \textbf{1.795} \\ 
        \midrule
        \textbf{Math} (106) & GNN & 41 & 60.583 & 69.5 & 82.25 & 8.452 & 9.302 \\ 
        \textbf{} & \toolwocctable & 46.923 & 71.077 & 82.615 & 94.538 & 2.922 & 3.297 \\ 
        \textbf{} & \tooltable & \textbf{55.25} & \textbf{82.917} & \textbf{91.917} & \textbf{99.417} & \textbf{1.748} & \textbf{2.103} \\ 
        \midrule
        \textbf{Mockito} (38) & GNN & 10.417 & 18 & 20.167 & 24.083 & 21.514 & 28.657 \\ 
        \textbf{} & \toolwocctable & 13.385 & 25.231 & 28.846 & 32.538 & 3.788 & 4.852 \\ 
        \textbf{} & \tooltable & \textbf{17.5} & \textbf{27.083} & \textbf{30.917} & \textbf{33.167} & \textbf{2.618} & \textbf{3.474} \\ 
        \midrule
        \textbf{Time} (26) & GNN & 7.5 & 11.75 & 14.333 & 17 & 19.587 & 22.058 \\ 
        \textbf{} & \toolwocctable & 10.769 & 14.692 & 16.154 & 18.385 & 10.893 & 11.752 \\ 
        \textbf{} & \tooltable & \textbf{11.083} & \textbf{15.333} & \textbf{17.417} & \textbf{20} & \textbf{8.948} & \textbf{10.034} \\ 
        \midrule
        \textbf{Total} (675) & GNN & 196.956 & 318.903 & 374.409 & 440.933 & 16.616 & 19.154 \\ 
        \textbf{} & \toolwocctable & 241.693 (23\%) & 375 (18\%) & 444.153 (19\%) & 520.077 (18\%) & 6.166 (63\%) & 7.285 (62\%) \\
        \textbf{} & \tooltable & \textbf{279.463 (42\%)} & \textbf{412.777 (29\%)} & \textbf{480.237 (28\%)} & \textbf{542.422 (23\%)} & \textbf{5.213 (68\%)} & \textbf{6.581 (65\%)} \\ 
        \bottomrule
    \end{tabular}
    }
    \vspace{-0.3cm}
\end{table}

\phead{Motivation.}  In real-world scenarios, a project might not have enough historical data to train a fault localization model. Hence, in this RQ, we explore the fault localization accuracy of \tool in a cross-project prediction scenario and compare the results against \textit{GNN}.

\phead{Approach.} We train \tool using the data from one specific project and apply the trained model to the remaining 13 projects. We repeat the process for every project. Then, for each project, we calculate the average fault localization accuracy of all the models trained using other projects. For example, to evaluate the cross-project result on Math, we train 13 models separately using other projects and apply the models to Math. We then calculate the average fault localization accuracy of these 13 models.  This ensures the models do not have any data leakage issues. We used \textit{GNN (i.e., Grace)} as the baseline and trained the models by following the same procedure. We only included \textit{GNN} in this RQ due to its better accuracy compared to other baselines as we found in RQ1. 

\phead{Results.} \textbf{\textit{Both \tool and \toolwocc outperform the baseline \textit{GNN} model in cross-project fault localization. Our models trained in cross-project settings even have higher or comparable accuracy compared to the baselines that were trained in same-project settings.}} Table \ref{tab:rq4} shows the cross-project fault localization results (i.e., cross-project setting). Across all the projects, \textit{DepGraph (cross-project)} achieved a Top-1 of 279.463, and \textit{\toolwocc (cross-project)} achieved a Top-1 of 241.693; both are noticeably better than the baseline \textit{GNN (cross-project)} that achieved a Top-1 of 196.956 (42\% and 23\% better, respectively). Particularly, \textit{DepGraph (cross-project)} showed an improvement of 23\% to 29\% in Top-3, Top-5, and Top-10, and over 65\% improvement in MFR and MAR, reflecting its higher fault localization accuracy. Although \textit{DepGraph (cross-project)} achieved better fault localization accuracy compared to \textit{\toolwocc (cross-project)}, \textit{\toolwocc (cross-project)} is still significantly better than \textit{GNN (cross-project)} across all evaluation metrics. The results highlight the importance of both our \graphname and code change information in improving GNN-based fault localization techniques.

Furthermore, \textit{DepGraph's} cross-project fault localization results are similar or even superior to that of other baselines that were trained and tested using data from the same project (i.e., same-project setting). Notably, as shown in RQ1, \textit{Ochiai}, and \textit{DeepFL} identified 121 and 257 faults in the Top-1 position, respectively, while \textit{DepGraph (cross-project) }achieved an average Top-1 of 279.463, better than \textit{Ochiai} by 131\% and \textit{DeepFL} by 8.7\%. We find that \textit{DepGraph (cross-project)} and \textit{GNN (same-project)} have similar fault localization results across all metrics, and \textit{DepGraph (cross-project)} even has better MFR and MAR scores. Our results show the potential of using \textit{DepGraph} in a cross-project setting due to its superior result even when compared to other fault localization techniques that are trained using data from the same project.

\rqboxc{
In cross-project settings, \tool achieves a 42\% higher Top-1 compared to \textit{GNN}. We also find that \tool in cross-project settings has better or comparable accuracy compared to other baselines trained in the same-project setting.}

\section{Threats to Validity}
\label{sec:threat}

\textit{Threats to internal validity} may arise from our technique implementations and experimental scripts. To address these, we have reviewed our code thoroughly and implemented it on state-of-the-art frameworks like Pytorch \cite{pytorch}. For \textit{Grace} and \textit{DeepFl}, we used the original implementations provided in prior work~\cite{lou2021boosting, li2019deepfl}. Another internal threat can be our manual analysis in the RQ3. The first two authors discussed each disagreement and analyzed the faults independently to mitigate subjectivity. 
\textit{Threats to external validity} may be tied to the benchmark used. We have countered this by testing on a popular benchmark Defects4J \cite{just2014defects4j}, featuring numerous real-world bugs and ensuring our techniques were evaluated on its latest version (V2.0.0). 
\textit{Threats to construct validity} may lie in the measurements of our study. To mitigate this, we adopted the leave-one-out cross-validation approach for a more generalizable study outcome which is also used in some prior studies \cite{lou2020can, lou2021boosting}. We also perform cross-project evaluations to mitigate data leakage issues. Our results show that even when trained in the cross-project scenario \tool still consistently outperforms the state-of-the-art GNN technique.

\section{Conclusion}
\label{sec:conclusion}

In this paper, we introduced a new graph neural network (GNN) based fault localization technique, \tool. \tool extends existing state-of-the-art GNN-based technique, {\it Grace}~\cite{lou2021boosting}, by developing \graphname, a unique graph representation that includes interprocedural call graph and software evolution details. 
Unlike previous graph representations, \graphnameLower removes edges between methods lacking call dependencies, thereby reducing graph noise and size. Afterward, we add code churn to the node attributes, providing historical code evolution information to the GNN.
Our evaluation on the Defects4j (V2.0.0) benchmark revealed that the \tool outperformed existing methods. Notably, \tool outperformed \textit{Grace}~\cite{lou2021boosting} across multiple metrics, highlighting the impact of our enhanced graph representation. \tool's enhanced graph structure, particularly when combined with code change data, showcased its effectiveness in identifying complex faults. Additionally, in cross-project settings, our approach demonstrated robust adaptability and significant performance improvements, underlining its potential in diverse software environments. Furthermore, with \graphname, we managed to reduce the GPU memory usage from 143GB to 80GB and speed up the training time from 9 days to just 1.5 days. Future research could explore integrating additional layers of data into the graph to further refine the accuracy of fault localization and optimize the needed resources for training/inference.


\section{Data Availability}
We have made our replication package available, which contains all the datasets and code available here: https://github.com/anonymoussubmission9/anonymous-submission \cite{AnonymousSubmission9}. 


\bibliographystyle{ACM-Reference-Format}
\bibliography{reference}

\end{document}